\documentclass[%
 reprint,
 amsmath,amssymb,
 aps,
]{revtex4-2}

\usepackage{graphicx}
\usepackage{bm}
\usepackage{comment}
\usepackage{caption}

\usepackage{xcolor}
\newcommand{\rev}[1]{#1}

\begin{document}


\title{\rev{Curvature}-Driven Density and Fluctuations in Interacting AOUPs under Confinement}
\thanks{A footnote to the article title}%
\author{Renu Parashar}
\author{Sandipan Dutta}%
\email{sandipan.dutta@pilani.bits-pilani.ac.in}
\affiliation{%
 Department of Physics, Birla Institute of Technology and Science, Pilani, Rajasthan, 333031, India 
}%

\date{\today}

\begin{abstract}

We investigate the steady-state dynamics of sterically interacting active Ornstein-Uhlenbeck particles confined in harmonic and double-well potentials, and compare their behavior with passive systems. By deriving a mean-field single-particle Fokker-Planck equation, we obtain effective density, force, diffusion, structure factor, and entropy-production descriptions that explicitly connect activity, interactions, and confinement geometry. \rev{
Correlation functions for harmonically confined degrees of freedom driven by
exponentially correlated active noise, including the normal modes of polymeric
systems, are well established
\cite{szamel2014self,samanta2016chain}.
Much less explored, however, is the regime in which excluded-volume interactions,
persistent active forcing, and spatially varying confinement curvature act
simultaneously. Here we address this regime analytically within mean-field,
small-persistence, and linearized approximations. We derive the stationary
single-particle density, velocity and position correlations, correlation length, dynamical structure factor, and leading-order entropy production for interacting
AOUPs under confinement. The collective correlations are expressed in terms of
the eigenvalues and eigenvectors of the configuration-dependent many-body Hessian, whose spectrum contains contributions from both the local curvature of
the external potential and interparticle interactions. This framework therefore
extends the familiar harmonic normal-mode description to an interacting system
with spatially nonuniform confinement. Two analytical results highlight the role of the confinement geometry.
First, within a locally homogeneous approximation, the correlation length is
$\xi(x^{\ast})=\sqrt{\tau K/
\big(\gamma+\tau\Phi''(x^{\ast})\big)}$,
showing that the spatial range of collective active fluctuations is controlled
jointly by persistence and the local curvature of the external potential.
The correlation length increases as the local curvature decreases and formally
diverges as
$\gamma+\tau\Phi''(x^{\ast})\to0^{+}$, signalling the loss of stability of
the locally linearized branch.
Second, at the leading nontrivial order retained in the small-$\tau$
description, the entropy production takes the form
$\dot S_{\rm tot}
\propto
\int dx\,\rho(x)
\left[
\partial_x^3\varphi_{\rm eff}(x)
\right]^2$,
linking irreversibility to spatial variations of the effective curvature.
The harmonic trap is therefore used as a constant-external-curvature reference,
while the double-well confinement provides a spatially nonuniform landscape in
which soft and stiff regions coexist. Their comparison reveals how persistence,
steric crowding, and local confinement geometry jointly control density
fluctuations, collective correlations, and irreversibility.
} In harmonic confinement, where the curvature is uniform, the system remains largely equilibrium-like, with smooth density profiles and weak entropy production. In contrast, the double-well potential separates the system into equilibrium-like regions near the well minima and nonequilibrium regions near the barrier. The barrier region exhibits enhanced velocity fluctuations, long-lived correlations, soft collective modes, increased density fluctuations, and larger entropy production. Persistence time amplifies these effects by increasing the localization of soft-modes near the barrier, while increasing particle number and steric interactions suppress fluctuations through crowding and drive the system toward an equilibrium-like state. The dynamical structure factor reveals enhanced low-frequency spectral weight from soft barrier modes, whereas entropy production is governed by spatial variations of the effective curvature. These results establish a direct link between confinement-induced curvature, active persistence, collective fluctuations, and irreversibility in interacting active matter.                
\end{abstract}

\maketitle


\section{\label{sec:introduction} Introduction}

Active matter systems are intrinsically out of equilibrium because their microscopic constituents continuously consume energy and convert it into \rev{self-propelled, temporally persistent motion. We emphasise at the outset that ``active'' here refers to this persistence and not to a net drift: the self-propulsion force has zero mean, and it is the finite correlation time of the propulsion -- equivalently the coloured, non-thermal character of the driving -- that breaks detailed balance and violates the fluctuation-dissipation relation.} This persistent energy injection breaks detailed balance and generates a wide range of collective phenomena, including flocking, clustering, giant number fluctuations, pattern formation, defect dynamics, anomalous transport, and collective flows in biological and synthetic systems \cite{marchetti2013hydrodynamics, bechinger2016active, shaik2023confined, ramaswamy2003active, ramaswamy2017active, prost2015active, doostmohammadi2018active, giomi2014defect, balasubramaniam2022active, balasubramaniam2022active, prost2015active, kruse2004asters, kruse2005generic,wang2021emergent}. A central goal of nonequilibrium statistical physics is to understand how microscopic activity modifies macroscopic observables such as density profiles, transport coefficients, fluctuations, correlations, collective modes, and entropy production.

Interacting active particles are of particular interest because interactions and persistence together produce collective behavior with no direct equilibrium analogue \cite{cates2012diffusive,bechinger2016active,ramaswamy2017active,fodor2018statistical,Touzo2025_Coulomb,Touzo2023_Dyson,Touzo2025_Riesz,fodor2016far}. A paradigmatic example is motility-induced phase separation, where purely repulsive active particles undergo phase separation in the absence of attractive forces \cite{fily2012athermal,cates2015motility,buttinoni2013dynamical,fily2014freezing,obyrne2023introduction}. Considerable progress has been made using hydrodynamic theories, coarse-grained descriptions, and perturbative approaches \cite{Martin2021,tailleur2008statistical,bialke2013microscopic,wittkowski2014scalar,solon2018generalized_njp,solon2018generalized_pre}. Nevertheless, an exact analytical description of interacting active matter remains difficult because activity introduces memory, nonlocal correlations, and irreversible probability currents. Exact or controlled results are therefore mostly available in special limits, such as two-particle systems \cite{slowman2016jamming,slowman2017exact,mallmin2019exact,das2020gap,metson2023from,metson2023tuning,ledoussal2021stationary,dolai2022inducing,hahn2023jamming,hahn2025steady}, harmonic chains \cite{Put2019,Singh2021,Prakash2024,Paul2024}, or specially constructed lattice models \cite{kourbane2018exact,Agranov2021,Agranov2023,Mukherjee2023,Thompson2011,Dandekar2020,Mukherjee2025}.

A particularly important and less understood regime is confined active matter, where external potentials break translational invariance and impose spatially varying curvature \cite{woillez2020active,woillez2020nonlocal,debnath2018activated,fily2019,caprini2021correlated,scacchi2019escape,dhar2019run,wexler2020dynamics,gu2020stochastic,woillez2020active,chaki2020escape,walter2021first,grandpre2018current,olsen2020escape,paoluzzi2020narrow,biswas2020first,schuss2007narrow,locatelli2015active,ghosh2014communication}. In such systems, transport, escape, fluctuations, and steady-state distributions are controlled not only by activity and interactions, but also by the local geometry of the trap. The interplay between persistence, crowding, and curvature can produce strongly heterogeneous nonequilibrium behavior that is absent in translationally invariant systems. Although confined active particles have been studied extensively, analytical theories for interacting confined systems remain limited because the dynamics depend sensitively on both many-body correlations and position-dependent curvature. Recent theoretical developments have begun to address this problem systematically \cite{marconi2015towards,maggi2015multidimensional,Touzo2026_DoubleWell}.

The double-well potential provides a natural setting to study these effects because it contains regions with qualitatively different local geometry within the same system. Near the well minima, the curvature is large and positive, and the dynamics is locally close to that in a harmonic trap. Near the barrier, the curvature becomes small or negative, restoring forces weaken, and persistent active motion can dominate relaxation. This spatial variation allows one to ask how activity reorganizes density, transport, fluctuations, correlations, and dissipation across different regions of a confined system. Caprini \textit{et al.}~\cite{caprini2019activejcp} and \rev{more recently Parashar et. al. \cite{Parashar2026_aoup}} demonstrated that the dynamics of an active particle in a double-well potential are strongly controlled by the confining landscape. Near the minima of the wells, termed Equilibrium Regions (ERs), the particle behaves approximately as an equilibrium system: the velocity distribution is nearly Gaussian, fluctuations are relatively small, mobility remains positive, and entropy production is weak. In contrast, near the barrier, where the curvature becomes small or negative, persistence dominates over the restoring force, leading to a qualitatively different dynamical regime. In this Nonequilibrium Region (NeR), the velocity distribution becomes bimodal, the mobility can become negative, entropy production is strongly enhanced, and barrier-crossing events are facilitated by persistent active motion. The coexistence of ER and NeR within the same potential illustrates how confinement geometry can spatially organize nonequilibrium behavior.

Motivated by the emergence of equilibrium-like and nonequilibrium regions in confined active systems, we develop a theoretical framework for interacting active Ornstein-Uhlenbeck particles (AOUPs) under confinement \cite{dabelow2019irreversibility, berthier2017new, mandal2017physical, Martin2021, nguyen2021active}. Using the perturbative steady-state approach of Martin \textit{et al.} \cite{Martin2021}, we derive a mean-field description for the single-particle distribution starting from the microscopic Langevin dynamics. Under the zero-current approximation, the active dynamics can be mapped onto an effective passive system characterized by a position-dependent diffusion coefficient, effective force, and effective potential. This framework allows us to obtain steady-state densities, transport coefficients, fluctuations, correlations, structure factors, and entropy production for interacting AOUPs in harmonic and double-well confinements.

We show that confinement geometry along with particle number and activity plays a central role in organizing nonequilibrium behavior. While harmonic traps produce smooth, nearly equilibrium-like density and fluctuation profiles, double-well confinement generates pronounced spatial heterogeneity. Persistence and particle number through excluded-volume interactions reshape the density distribution and renormalize the effective transport properties. By linearizing the dynamics around the steady-state configuration, we derive analytical expressions for velocity and position correlations in terms of the Hessian spectrum of the effective many-body potential. The local curvature controls the crossover between equilibrium-like regions (ERs), where stiff modes suppress fluctuations and correlations, and nonequilibrium regions (NeRs), where soft barrier modes amplify fluctuations, increase correlation lengths, and promote collective barrier-crossing events. These soft modes also dominate the dynamical structure factor, producing enhanced low-frequency spectral weight and persistence-induced memory effects. Increasing the particle number increases the stiffness of the Hessian eigenmodes, leading to reduced fluctuations. Finally, we show that entropy production is directly governed by spatial variations of the effective curvature, leading to weak dissipation in harmonic confinement but strong localized irreversibility near the barrier and inflection regions of the double-well potential. We investigate how confinement geometry spatially organizes nonequilibrium behavior: stiff, positively curved regions behave approximately as equilibrium-like regions, whereas soft barrier regions act as nonequilibrium regions with enhanced fluctuations, long-lived correlations, collective crossings, and localized entropy production. This establishes curvature and particle number as a central organizing principle for interacting active matter under confinement.

\rev{\textbf{We now state explicitly the novelty of the present work and the gap that it addresses.}
Correlation functions for a harmonically confined degree of freedom driven by exponentially correlated active noise are well established
\cite{szamel2014self,samanta2016chain}.
In particular, Samanta and Chakrabarti extended the harmonic active-noise calculation to the normal modes of Rouse, Zimm, and semiflexible polymer chains
\cite{samanta2016chain}, where each mode obeys an effective linear harmonic equation.
The spatial organization of equilibrium-like and nonequilibrium regions in a double-well potential has also been studied for a single active particle
\cite{caprini2019activejcp}.
Analytical treatments of interacting active systems have therefore covered several important limiting cases, including single-particle confinement, linear or harmonic normal-mode systems, translationally invariant interacting systems, and few-particle models.
Much less explored is the regime in which excluded-volume interactions, persistent active forcing, and a spatially varying external curvature act simultaneously.
The present work addresses this regime and develops the following analytical results within the mean-field, small-persistence, and linearized approximations specified below.}

\begin{itemize}

    \item \rev{
    \textbf{Mean-field theory for interacting AOUPs in a general external confinement.}
    Starting from the many-particle dynamics, we derive a closed single-particle density equation from the BBGKY hierarchy using a two-body factorisation closure, Eq.~\eqref{bbgkymeanfield}.
    This leads to an effective description in terms of the position-dependent diffusivity
    $D_{\rm eff}(x)$, the effective force $\mathcal F(x)$, and the effective potential
    $U_{\rm eff}(x)$, Eqs.~\eqref{density}--\eqref{deff},
    \eqref{interactingdensity}, and \eqref{effectivepotential}.
    In these expressions, interparticle interactions and the spatial structure of the external confinement enter simultaneously.
    The corresponding non-interacting UCNA expression is recovered as the limiting case
    $V\rightarrow 0$, Eq.~\eqref{noninteractingdensity}.
    }

    \item \rev{
    \textbf{Collective velocity and position correlations from the interacting many-body Hessian.}
    The velocity and position correlations of the confined interacting system,
    Eqs.~\eqref{velocitycorrelations}--\eqref{fluctuation}, are expressed in terms of the eigenvalues
    $\lambda_\alpha$ and eigenvectors $\psi_\alpha$ of the many-body Hessian,
    Eq.~\eqref{hessianmatrix}.
    This should be distinguished from the earlier linear harmonic-mode calculations
    \cite{szamel2014self,samanta2016chain}.
    In those problems, the relevant degrees of freedom can be represented by independent harmonic modes with prescribed mode stiffnesses.
    In the present system, by contrast, the fluctuation operator is
    configuration dependent and contains both the local curvature of the external potential and the stiffness generated by interparticle interactions,
$H_{ij}=\Phi''(x_i^\ast)\delta_{ij}+\partial_{x_i}\partial_{x_j}V(x_i^\ast-x_j^\ast)$.
    The resulting spectrum therefore depends simultaneously on confinement geometry and steric coupling.
    The corresponding harmonic linear-mode expressions are recovered in the appropriate limiting case and are used here as a consistency check rather than as a new result.
    The eigenvectors $\psi_\alpha$ additionally provide the spatial structure of the collective modes and allow regions of enhanced fluctuations and soft collective response to be identified inside a spatially nonuniform confinement.
    }

    \item \rev{
    \textbf{A position-dependent correlation length controlled by local confinement curvature.}
    Within a locally homogeneous approximation,
$\lambda(q;x^\ast)\simeq\Phi''(x^\ast)+Kq^2$,
    the correlation function defines the local correlation length $ \xi(x^\ast)=\sqrt{\tau K/\big(\gamma+\tau\Phi''(x^\ast)\big)}$,
    showing explicitly that the spatial range of collective active fluctuations is controlled jointly by persistence and the local curvature of the confinement.
    As the local curvature decreases, $\xi(x^\ast)$ increases.
    A formal divergence occurs when $\gamma+\tau\Phi''(x^\ast)\rightarrow 0^{+}$,
    corresponding to the loss of stability of the locally linearized branch.
    For a harmonic external potential, $\Phi''(x)=k$ is spatially constant, and therefore no position dependence of $\xi$ originates from the external confinement itself.
    }

    \item \rev{
    \textbf{Entropy production associated with spatial variation of the effective curvature.}
    Within the stated small-persistence approximation, the leading nontrivial contribution to the entropy production is
        $\dot S_{\rm tot}\propto\int dx\, \rho(x)\left[\partial_x^3\varphi_{\rm eff}(x)
        \right]^2$ 
    which relates irreversibility to the spatial variation of the effective curvature.
    A non-interacting harmonic potential has constant curvature and therefore gives no contribution at the order retained here.
    Interactions can nevertheless generate a finite contribution through the interaction-dependent part of
    $\varphi_{\rm eff}$.
    For a nonlinear and spatially nonuniform confinement such as the double well, the external potential itself also contributes because its curvature varies with position.
    }

    \item \rev{
    \textbf{Competition between persistence and crowding.}
    Within the parameter regime investigated, increasing the persistence time $\tau$ enhances the contribution of soft collective modes, particularly those associated with the barrier region, and increases the spatial heterogeneity of the nonequilibrium response.
    Increasing particle number or steric interaction strength, on the other hand, generally stiffens the relevant collective modes and suppresses the corresponding fluctuations.
    Persistence, confinement curvature, and crowding therefore act as competing controls of the collective nonequilibrium response.
    The simulations are used to test and illustrate these analytical predictions rather than serving as the sole basis for the conclusions.
    }

\end{itemize}

\rev{\textbf{We also clarify the role of the harmonic trap in the present study.}
The harmonic problem and related linear normal-mode systems are well studied
\cite{szamel2014self,samanta2016chain}, and we do not present the harmonic trap itself as a new active-matter problem.
Instead, it is used as a constant-external-curvature reference system, for which
$\Phi''(x)=k$, and $\Phi'''(x)=0$ everywhere.
Consequently, the external potential introduces no spatial variation of curvature.
For an interacting system, however, the many-body Hessian can still possess a nontrivial spectrum because of the interaction contribution.
The harmonic confinement therefore provides a useful reference for separating interaction-induced collective effects from effects generated specifically by a spatially varying external potential.}

\rev{
The double-well potential provides the complementary case in which the external curvature changes strongly with position.
It is positive and relatively large around the well minima, decreases on approaching the barrier, and becomes negative in the barrier region.
Consequently, different parts of the same confined system sample locally stiff, soft, and locally unstable curvatures.
Comparing harmonic and double-well confinements at fixed persistence, interaction strength, and particle number therefore provides a controlled way of identifying the additional effects produced by spatially nonuniform confinement and soft barrier modes.
This comparison motivates our central analysis of how local confinement geometry, persistent active forcing, and crowding jointly control density profiles, fluctuations, collective correlations, soft modes, and irreversibility.
}

The paper is organized as follows.
 In Section~\ref{sec:aoup}, we introduce the Langevin dynamics of interacting AOUPs and derive the corresponding mean-field Fokker-Planck equation. In Section~\ref{sec:steadystatedensities}, we obtain the steady-state density, effective diffusion constant, and effective force for harmonic and double-well confinements. In Section~\ref{sec:steadystatevelocities}, we analyze velocity and position fluctuations by linearizing the dynamics around the steady-state particle positions. In Section~\ref{sec:structurefactor}, we derive the static and dynamical structure factors and compare passive and active responses. In Section~\ref{sec:entropyproduction}, we compute the entropy production rate and relate it to curvature variations, particle distributions, and nonequilibrium fluctuations.

\section{\rev{Dynamics of AOUP and passive particles in a confining potential}}
\label{sec:aoup}
Consider a one-dimensional system of $N$ AOUPs interacting through a potential $V(x_i-x_j)$ where $\{x_i\}_{i=1,2,...,N}$ are the positions of the particles. They diffuse in a confining potential $\Phi(x_i)$. \rev{Throughout this section $\Phi(x)$ is an arbitrary confining potential; the harmonic and double-well forms, Eqs.~\eqref{harmonic_trap} and \eqref{double_well}, are specified only from Sec.~\ref{sec:steadystatedensities} onwards.} Their dynamics is governed by
\begin{align}
\gamma\dot{x}_i & = -\Phi^{\prime}(x_i) + F_i +\eta_i.
\label{aoup}
\end{align}
$\eta_i$ is an Ornstein-Uhlenbeck process : $\langle\eta_i(t)\eta_j(t^{\prime})\rangle=\frac{v_0^2}{2}\exp(-\vert t-t^{\prime}\vert/\tau)$ with persistence time $\tau$ and variance $v_0$. The active diffusion constant is given by $D = v_0^2\tau$. $\gamma$ is the friction coefficient. The time evolution of the Ornstein-Uhlenbeck noise can also be written as
\begin{equation}
\dot{\eta}_i = -\frac{1}{\tau}\eta_i+\frac{\sqrt{2D}}{\tau}\xi_i,
\label{ou-noise}
\end{equation}
where $\langle\xi_i(t)\xi_j(t^{\prime})\rangle=\delta_{ij}\delta(t-t^{\prime})$.

\rev{
Here $\eta_i$ has zero mean by construction, $\langle \eta_i\rangle=0$, while its finite correlation time $\tau$
introduces persistent nonequilibrium forcing. A vanishing mean particle
velocity is not, by itself, a signature of passivity: in a confined stationary state, standard active-particle models such as active Brownian
particles and run-and-tumble particles also satisfy
$\langle \dot{x}\rangle=0$, since there is no sustained translational current in the confining direction. The term "active" therefore refers
here not to a nonzero mean drift, but to the presence of a persistent
self-propulsion whose statistics are not fixed by the equilibrium
fluctuation--dissipation relation
\cite{fodor2016far,szamel2014self}. With our convention,
$\langle \eta_i(t)\eta_j(t')\rangle
=(v_0^2/2)\delta_{ij}\exp[-|t-t'|/\tau]$, the parameter $v_0$
sets the root-mean-square amplitude of the active propulsion rather than
a mean drift velocity. For generic nonlinear confinement or interacting
systems, this persistent driving produces a nonequilibrium steady state.
The harmonic potential is an important special case: because its
curvature is constant, the AOUP dynamics can remain equilibrium-like and
the leading small-$\tau$ entropy production vanishes. Nonlinear
confinement, such as the double-well potential considered here, removes
this special property and allows persistent driving to generate spatially
heterogeneous nonequilibrium behaviour. In the white-noise limit
$\tau\to0$ at fixed active diffusivity $D$, the Ornstein--Uhlenbeck
propulsion reduces to an effective thermal-like white noise and the
passive Brownian limit is recovered.
}

\rev{Thermal fluctuations can be included by adding an independent Gaussian
white-noise term $\gamma\sqrt{2D_T}\,\zeta_i(t)$ to Eq.~\eqref{aoup}, with
$\langle \zeta_i(t)\zeta_j(t') \rangle=\delta_{ij}\delta(t-t')$ with
$D_T=\frac{k_B T}{\gamma}$.
At the level of the coarse-grained density equation, this produces the
usual translational-diffusion term $D_T\partial_x^2\rho$ and therefore
adds a thermal contribution to the effective diffusivity. Thermal noise
also contributes the standard equilibrium fluctuations; for example,
in the harmonic approximation the equal-time positional variance of a
normal mode $\alpha$ acquires the equilibrium contribution
$k_B T/\lambda_\alpha$. We set $D_T=0$ here in order to isolate the effects of persistent active
forcing. Since thermal noise satisfies the equilibrium
fluctuation--dissipation relation, it is not itself a source of
nonequilibrium driving. Its inclusion would primarily smooth the
steady-state density profiles and reduce the contrast of
activity-induced spatial heterogeneities. The results presented below
should therefore be understood primarily in the activity-dominated
regime.}

$F_i = -\partial_{x_i}\left(\sum_{j<i}V(x_i-x_j)\right)$ is a conservative force due to the interactions between the particles. Following \cite{caprini2018active}, we rewrite the entire velocity term on the right-hand side of Eq.\eqref{aoup} as $\nu_i = \frac{1}{\gamma}[-\Phi^{\prime}(x_i) + F_i +\eta_i]$, which has a conservative part originating from the external or interaction potential and a dissipative part due to self-propulsion. Eq.\eqref{aoup} and Eq.\eqref{ou-noise} become
\begin{widetext}
\begin{align}
\dot{x}_i & = \nu_i \nonumber\\
\dot{\nu}_i & = -\sum_j\left[\frac{1}{\tau}\delta_{ij}+\frac{1}{\gamma}\Phi^{\prime\prime}(x_i)\delta_{ij}+\frac{1}{\gamma}\partial_{x_ix_j}V(x_i-x_j)\right]\nu_j-\frac{1}{\gamma\tau}\Phi^{\prime}(x_i)+\frac{1}{\gamma\tau}{F_i}+\frac{\sqrt{2D}}{\tau}\xi_i.
\label{langevin}
\end{align}
\end{widetext}

 Integrating the multiparticle Fokker-Planck equation over the positions and velocities of all particles except one in Appendix \ref{Appendix:Fokker-Planck}, we obtain the single particle probability distribution in the mean-field approximation 
\begin{widetext}
\begin{align}
\dot{P}_1(x,\nu, t) &  = -\partial_x[\nu P_1(x,\nu, t)]+\partial_{\nu}\biggl[\frac{1}{\tau}\nu P_1(x,\nu, t)\biggr]+\partial_{\nu}\biggl[\int dx^{\prime}d\nu^{\prime}\frac{1}{\gamma}\partial_{xx^{\prime}}V(x-x^{\prime})\nu^{\prime}P_1(x,\nu, t) \nonumber \\ & P_1(x^{\prime},\nu^{\prime}, t)\biggr]  +\partial_{\nu}\biggl[\frac{1}{\gamma}\nu \partial_{x}^2\Phi(x)P_1(x,\nu, t)\biggr] +\frac{1}{\gamma\tau}{\partial_x\Phi(x)}\partial_{\nu}P_1(x,\nu, t)+\frac{1}{\gamma\tau}\partial_{\nu}P_1(x,\nu, t)\times \nonumber \\ &\int dx^{\prime}d\nu^{\prime} {\partial_xV(x-x^{\prime})}P_1(x^{\prime},\nu^{\prime}, t)+\frac{D}{\tau^2}\partial_{\nu}^2P_1(x,\nu, t).
\label{bbgkymeanfield}
\end{align}
\end{widetext}

 The evolution equation for the density of particles $\rho(x,t) = \int d\nu P_1(x,\nu,t)$ is
\begin{widetext}
\begin{align}
\partial_t\rho(x,t) & = \partial_x j(x) =  \partial_x\biggl[D_{eff}(x)\partial_x\rho(x,t)+\mathcal{F}(x)\rho(x,t)\biggr].
\label{density}
\end{align}
\end{widetext}
The detailed derivation is provided in the Appendix \ref{Appendix:Fokker-Planck}. We have defined 
\begin{widetext}
\begin{align}
D_{\text{eff}}(x) & = \frac{D}{\left(1+\frac{\tau}{\gamma}\Phi^{\prime\prime}(x)\right)\Gamma(x)},\text{ and} \label{deff} \\
\mathcal{F}(x) & = \frac{1}{\Gamma(x)}\left[\frac{1}{\gamma}\Phi^{\prime}(x) + \frac{1}{\gamma}\int dx^{\prime}\rho(x^{\prime})\partial_xV(x-x^{\prime}) + D\partial_x\left[\frac{1}{1+\frac{\tau}{\gamma}\Phi^{\prime\prime}(x)}\right]\right]. 
\end{align}
\end{widetext}

\rev{
It is important to state explicitly which approximations enter
Eqs.~\eqref{bbgkymeanfield}--\eqref{deff}, and in what sense the
description is a mean-field one.
(a) \textit{Closure of the BBGKY hierarchy.}
The exact first hierarchy equation, Eq.~\eqref{bbgky}, couples $P_1$ to
the two-body distribution $P_2$. We close it by the factorisation
\[
P_2(x,\nu,x',\nu',t)
=
P_1(x,\nu,t)P_1(x',\nu',t),
\]
i.e.\ we neglect two-body correlations in the interaction term and replace
the force on a tagged particle by the force generated by the average
one-body density of the remaining particles. This is the only sense in
which the term ``mean field'' is used in the present work. The closure is
expected to be more reliable when correlations are weak, when each particle
samples many neighbours, or when the density varies slowly over the
interaction range. It is not a controlled approximation for strongly
correlated, dilute, or strongly ordered configurations. Since the present
interactions are short-ranged and steric, we therefore regard this
factorisation as an approximation whose accuracy must ultimately be assessed
against the microscopic dynamics.
(b) \textit{Truncation of the velocity-moment hierarchy.}
Equations~\eqref{current}--\eqref{susceptibility} for $j$ and
$\mathcal V$ involve the third moment $\chi$. We assume that
$\mathcal V(x,t)$ varies slowly on time scales long compared with $\tau$
and that $\partial_x\chi\simeq 0$. This is the standard adiabatic
(local-equilibrium) truncation, valid when the velocity distribution
relaxes rapidly compared with the density
\cite{Martin2021}.
(c) \textit{Neglect of the current gradient.}
In deriving Eq.~\eqref{appendix:current} we neglect $\partial_x j$,
which is justified in and near the steady state when the current is small
and slowly varying. In Sec.~\ref{sec:steadystatedensities} we impose
$j=0$ exactly for the steady-state solutions.
(d) \textit{Small-persistence expansion.}
Equations~\eqref{deff} and \eqref{entropy} are obtained to leading
non-trivial order in $\tau$, with corrections of
$\mathcal O(\tau^3)$. The corresponding coarse-grained expressions are
therefore controlled only when the dimensionless persistence correction,
for example $\tau\Phi''(x)/\gamma$, remains sufficiently small over the
region in which the approximation is applied.
(e) \textit{Smooth-density approximation.}
The approximation $\rho''''(x)\approx 0$ is used only in the simplified
estimate of $D_{\rm eff}(x)$ quoted below Eq.~\eqref{eq:effective}; it is
not used in generating the figures.
}

\rev{
A further condition must be satisfied for the effective passive
description to remain physically meaningful. From Eq.~\eqref{deff},
$D_{\rm eff}(x)
=
\frac{D}{
\left(1+\frac{\tau}{\gamma}\Phi''(x)\right)\Gamma(x)
}$,
and we require
$1+\frac{\tau}{\gamma}\Phi''(x)>0$,
and
$\Gamma(x)>0$,
throughout the spatial region sampled by the particles. Under these
conditions, $D_{\rm eff}(x)$ is positive. The first condition is the
familiar positivity and invertibility requirement associated with
UCNA-type coarse graining. It can fail in regions of sufficiently
negative external curvature,
$\Phi''(x)<-\frac{\gamma}{\tau}$,
which in a double-well potential may occur near the barrier at sufficiently
large persistence. The second condition imposes an additional restriction
through the interaction contribution,
$\Gamma(x)
=
1+\frac{\tau}{\gamma}\Phi''(x)-\tau\Psi(x)$.
The constraint arising from the external potential alone may be written as $\tau<
\frac{\gamma}{\left|\min_x\Phi''(x)\right|}$. For the double-well potential of Eq.~\eqref{double_well},
$\Phi(x)=\frac{k}{4}(x^2-4)^2$,
one has
$\Phi''(x)=k(3x^2-4),
\qquad
\min_x\Phi''(x)=\Phi''(0)=-4k$,
so that the corresponding bare-curvature bound is
$\tau<\frac{\gamma}{4k}$.
In addition, the interaction-dependent condition $\Gamma(x)>0$ must be
verified separately. For the parameter values used in
Figs.~\ref{fig:ssdensity}--\ref{fig:individual_particles}, both conditions
are satisfied and $D_{\rm eff}(x)$ remains positive throughout
Fig.~\ref{fig:effective_system}. A violation of either condition should not be interpreted as a physical
negative diffusivity; rather, it signals the breakdown of the
coarse-grained effective-passive description, in which case the full
velocity-resolved dynamics of Eq.~\eqref{bbgkymeanfield} must be retained.
}

\section{Steady-state densities and Effective forces}
\label{sec:steadystatedensities}

The steady-state solution in the zero current approximation $j = 0$ in Eq.\eqref{appendix:current} has the following solution
\begin{widetext}
\begin{align}
\rho(x) & =\mathcal{N}\exp\biggl(-\frac{1}{D\gamma}\biggl[\Phi(x)+\frac{\tau}{2\gamma}(\Phi^{\prime}(x))^2-D\gamma\ln\left(1+\frac{\tau}{\gamma}\Phi^{\prime\prime}(x)\right)+\int dx^{\prime}V(x-x^{\prime})\rho(x^{\prime})\nonumber\\&+\frac{\tau}{\gamma}\int\Phi^{\prime\prime}(x^{\prime})\int^xV^{\prime}(x^{\prime}-x^{\prime\prime})\rho(x^{\prime\prime})\biggr]\biggr).
\label{interactingdensity}
\end{align}
\end{widetext}
$\mathcal{N}$ is the normalization factor determined from $\int dx\rho(x) = N$.

In the non-interacting case setting $V = 0$ in Eq.\eqref{interactingdensity} immediately leads to
\begin{widetext}
\begin{equation}
\ln\rho(x) = -\frac{1}{D\gamma}\left[\Phi+\frac{\tau}{2\gamma}(\Phi^{\prime})^2-D\gamma\ln[1+\frac{\tau}{\gamma}\Phi^{\prime\prime}]\right] + \text{const}.
\label{noninteractingdensity}
\end{equation}
\end{widetext}
The right-hand side of the equation is the effective potential obtained within the UCNA approximation in \cite{jung1987dynamical, maggi2015multidimensional}.

Eqn \eqref{density} can be mapped to a passive system \cite{tailleur2008statistical} with dynamics
\begin{equation}
 \dot{x}_i = F_{\text{eff}}(x_i) +\sqrt{2D_{\text{eff}}(x_i)}\epsilon_i(t).
 \label{effectivelangevin}
\end{equation}
$\{\epsilon_i\}$ are Gaussian white noises with unit variances. The effective force acting on each particle is
\begin{equation}
F_{\text{eff}}(x) = -\mathcal{F}(x)+\partial_{x}D_{eff}(x)
\label{effectiveforce}
\end{equation}

The effective potential experienced by individual particles can be defined by the term in the exponent of the density in Eq.\eqref{interactingdensity}
\begin{widetext}
\begin{align}
U_{\text{eff}}(x) & =  \Phi(x)+\frac{\tau}{2\gamma}(\Phi^{\prime}(x))^2-D\gamma\ln\left(1+\frac{\tau}{\gamma}\Phi^{\prime\prime}(x)\right)+\int dx^{\prime}V(x-x^{\prime})\rho(x^{\prime})\nonumber\\&+\frac{\tau}{\gamma}\int\Phi^{\prime\prime}(x^{\prime})\int^xV^{\prime}(x^{\prime}-x^{\prime\prime})\rho(x^{\prime\prime}).
\label{effectivepotential}
\end{align}
\end{widetext}

\begin{figure*}[!tbh]
  \includegraphics[width=\linewidth]{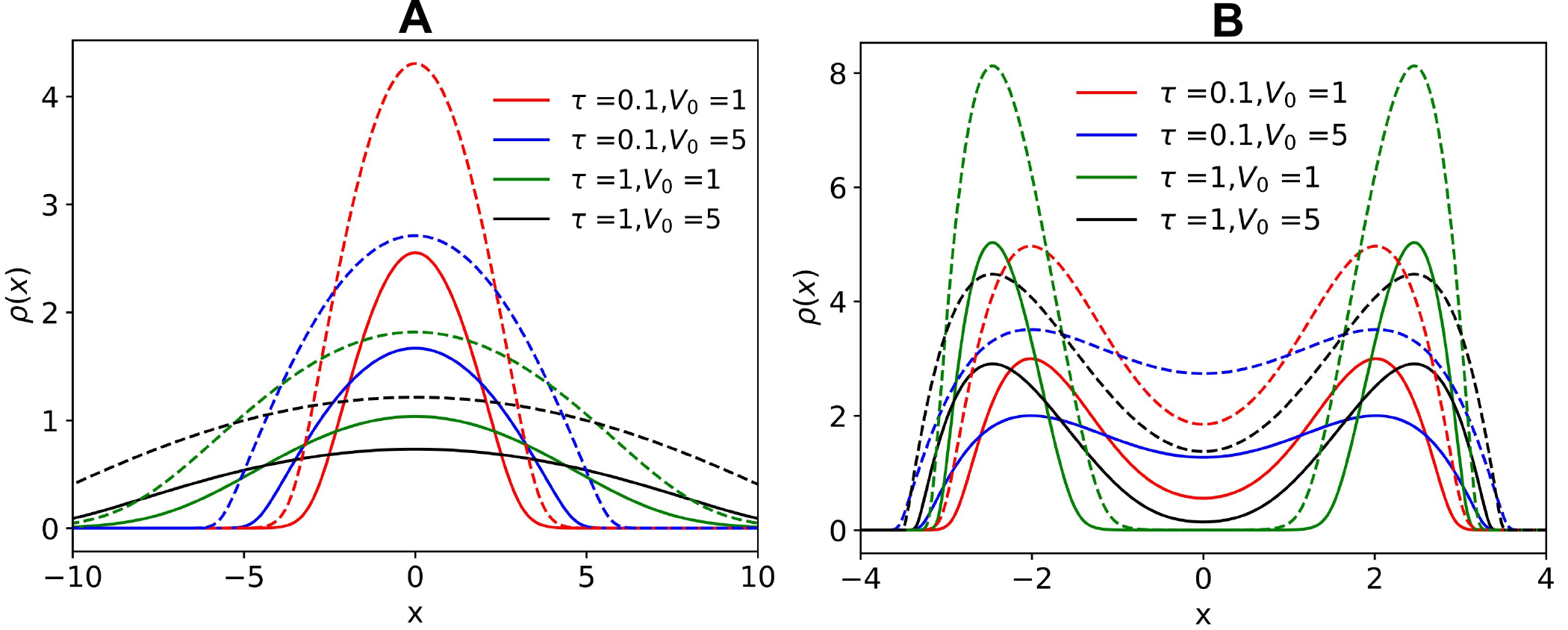}
  \caption{Steady-state density profiles of interacting AOUPs under confinement. (A) Density distribution in a harmonic trap, with enhanced accumulation and broadening controlled by the interaction strength $V_0$, particle number $N$, and persistence time $\tau$. The solid (dashed) curves represent $N = 10$ ($N = 15$) particles.  (B) Density distribution in a double-well potential, illustrating activity-driven localization in the wells and modified barrier occupancy, where increasing $V_0$, $N$, and $\tau$ amplifies crowding effects and sharpens nonequilibrium density contrasts between the wells and the barrier region. The solid (dashed) curves represent $N = 10$ ($N = 15$) particles.}
  \label{fig:ssdensity}
\end{figure*}

\begin{figure*}
  \includegraphics[width=\linewidth]{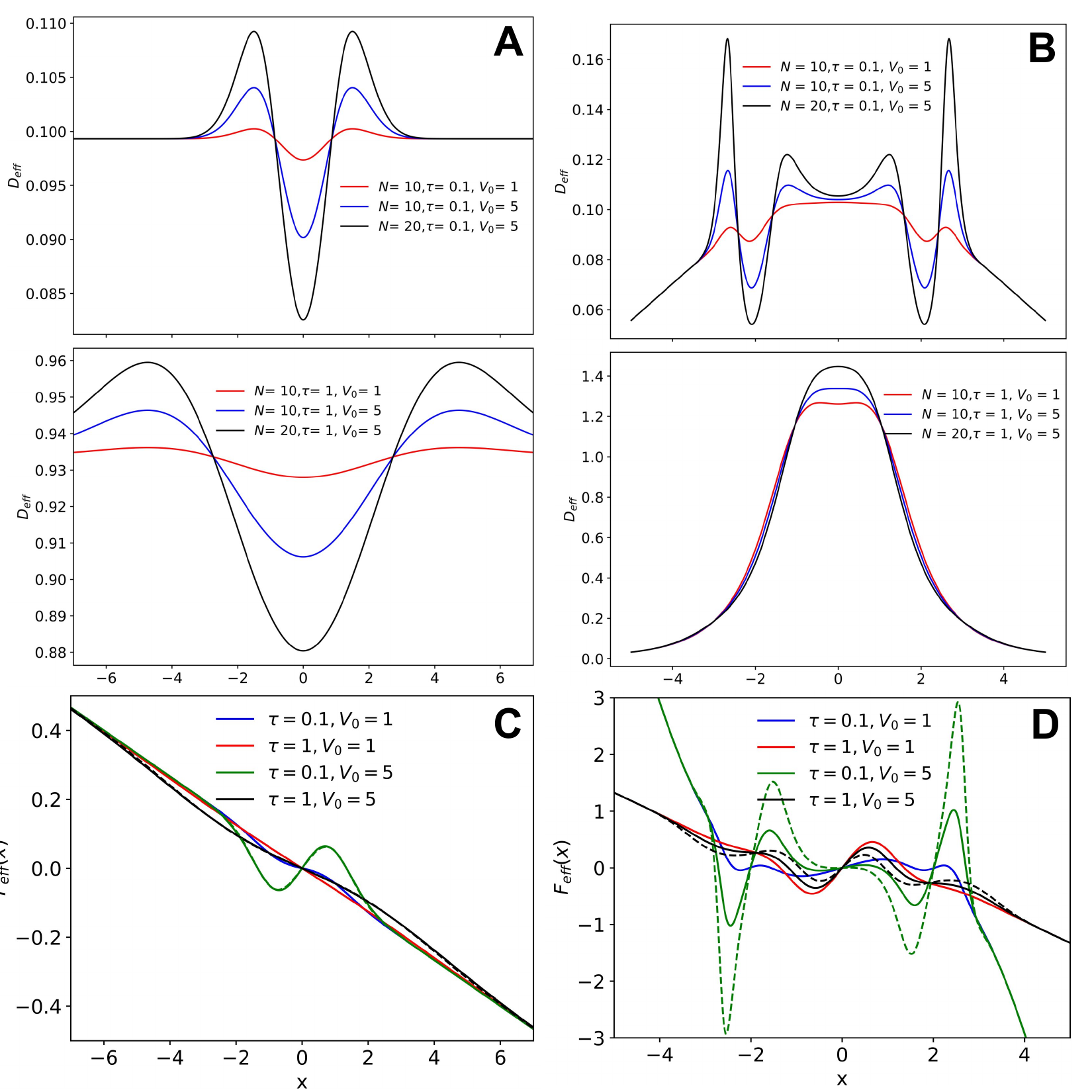}
  \caption{ Local transport and effective force in interacting AOUPs under confinement.
(A) Spatially resolved local diffusion constant in a harmonic trap for two persistence times, $\tau = 0.1$ and $\tau = 1$, showing interaction- and density-induced suppression and enhanced heterogeneity with increasing WCA strength $V_0$ and particle number $N$.
(B) Local diffusion constant in a double-well potential, highlighting strong spatial variation near the well minima and barrier, with reduced diffusion at larger $V_0$, $N$, and $\tau$.
(C) Effective force experienced by a particle in a harmonic trap, illustrating activity- and interaction-driven renormalization of the trap stiffness.
(D) Effective force in a double-well potential, exhibiting pronounced nonlinearity and barrier reshaping due to the combined effects of persistence time $\tau$, excluded-volume interactions $V_0$, and crowding $N$.}
  \label{fig:effective_system}
\end{figure*}

We consider $N$ AOUPs interacting via short-ranged potential in two kinds of confinements: a harmonic trap 
\begin{equation}
 \Phi(x) = \frac{k}{2}x^2
 \label{harmonic_trap}
\end{equation}
 centered at the origin, and a double-well potential 
 \begin{equation}
 \Phi(x) = \frac{k}{4}(x^2-2^2)^2 
 \label{double_well}
 \end{equation}
 with wells located at $x =\pm 2$. \rev{
These two confinements are studied together for a specific reason. The
harmonic trap is not presented here as a new problem---it is very well
studied---but is used as a constant-external-curvature reference system,
for which
$\Phi''(x)=k$,
and
$\Phi'''(x)=0$
everywhere. Thus, the external confinement itself introduces no spatial
variation of curvature. In the interacting harmonic system, however, the
many-body Hessian can still possess a nontrivial spectrum because of the
interaction contribution. The harmonic trap therefore provides a useful
reference in which interaction-induced spatial heterogeneity can be distinguished from that generated by spatial variations of the external
confinement.
The double-well potential of Eq.~\eqref{double_well}, by contrast, has a
strongly position-dependent curvature. For
this,
one has
$\Phi''(x)=k(3x^2-4)$,
so that
$\Phi''(\pm2)=8k$,
and
$\Phi''(0)=-4k$.
The well minima therefore correspond to regions of large positive
curvature, whereas the barrier is a region of negative curvature.
Comparing the harmonic and double-well confinements at fixed persistence
time $\tau$, interaction strength $V_0$, and particle number $N$ provides
a controlled way of identifying the additional effects associated with
spatially nonuniform external confinement and the presence of soft barrier
regions. This comparison is particularly useful for interpreting the density,
local diffusivity, fluctuation spectrum, correlation length, dynamical
structure factor, and entropy production. In particular, because
$\Phi'''(x)=0$ in the harmonic trap, any contribution to the
leading-order entropy production arising there must originate from the
interaction-dependent part of the effective potential, whereas in the
double-well confinement the spatial variation of the external curvature
provides an additional contribution to irreversibility.
} First we focus on harmonic confinement. For short ranged potential $V$, the integral $\int dx^{\prime}V(x-x^{\prime})\rho(x^{\prime}) = V_0\rho(x) + V_1\rho^{\prime\prime}(x)$ \cite{chavanis2008hamiltonian} in Eq. \eqref{interactingdensity}, the density simplifies to 
\begin{align}
\ln\rho(x)\propto -\frac{k+\frac{k^2\tau}{\gamma}}{2D\gamma}x^2-\left(1+\frac{k\tau}{\gamma}\right)\left(V_0\rho(x)+V_1\rho^{\prime\prime}(x)\right).
\label{eq:harmonicdensity}
\end{align}

The steady-state density profile of an AOUP reflects the interplay between confinement, persistence time $\tau$, and interparticle repulsion controlled by the WCA strength $V_0$. In a single-well harmonic potential (Fig.\ref{fig:ssdensity}-(A)), the density remains unimodal and centered at the trap minimum, but it deviates from a simple Gaussian as $V_0$ increases. For small 
$\tau$ (low activity), the AOUP behaves close to an effective equilibrium particle, yielding a narrow, nearly Boltzmann-like profile. As 
$\tau$ increases, persistent self-propulsion pushes particles outward, broadening the distribution and enhancing the density near the trap edges; strong repulsive interactions ($V_0$ large) further flatten the central peak by suppressing local crowding, leading to a wider and more uniform density around the center. The double-well trap (Fig.\ref{fig:ssdensity}-(B)) produces a bimodal density with peaks near the minima of each well. At low activity ($\tau$), the density is strongly localized in wells with weak interwell population due to rare crossings. Increasing $\tau$ improves persistence-driven barrier crossing, redistributing probability towards the barrier region, and reducing depth contrast between wells. Stronger repulsive interactions amplify this effect by penalizing accumulation at the minima, promoting broader peaks and, at large $\tau$, a partially filled barrier region, signaling a strongly nonequilibrium steady-state where activity effectively lowers the barrier and reshapes the density far from equilibrium expectations.

The effective force and effective diffusion constant of the effective system in a harmonic trap (Eq.\eqref{harmonic_trap}) reads
\begin{widetext}
\begin{align}
D_{\text{eff}}(x) & = \frac{D}{1+\frac{k\tau}{\gamma}+\frac{k^2\tau^2}{\gamma^2}-\frac{\tau}{\gamma}\left(V_0\rho^{\prime\prime}(x)+V_1\rho^{\prime\prime\prime\prime}(x)\right)}, \text{ and } \nonumber \\ \mathcal{F}(x) & = \frac{\left(1+\frac{k\tau}{\gamma}\right)\left(\frac{k}{\gamma}x+\frac{V_0}{\gamma}\rho^{\prime}(x)\right)}{1+\frac{k\tau}{\gamma}+\frac{k^2\tau^2}{\gamma^2}-\frac{\tau}{\gamma}\left(V_0\rho^{\prime\prime}(x)+V_1\rho^{\prime\prime\prime\prime}(x)\right)}.
\label{eq:effective}
\end{align}
\end{widetext}
Assuming sufficiently smooth densities such that $\rho^{\prime\prime\prime\prime}(x)\approx 0$ and using the non-interacting density (implying $\rho^{\prime\prime}(x)\approx -V_0\rho(x)/V_1$ from Eq.\eqref{eq:harmonicdensity}), we get $D_{\text{eff}}(x) \approx \frac{v_0^2\tau}{1+\frac{k\tau}{\gamma}+\frac{k^2\tau^2}{\gamma^2}+\frac{\tau}{\gamma}V_0^2\rho(x)}$. The local diffusion constant depends non-monotonically on $\tau$ reaching maximum when it equals the relaxation time in the trap $\gamma/k$. The excluded-volume interactions strongly suppress diffusion in a density-dependent manner: so regions of high local density exhibit a markedly reduced $D_{eff}$.  Overall, larger $V_0$ and $N$ promote strong local suppression of diffusion via crowding, while increasing $\tau$ enhances nonequilibrium correlations that compete with and magnify confinement and interaction effects, resulting in pronounced spatial variation of the local diffusion constant. Fig.\ref{fig:effective_system} summarizes how activity and interactions renormalize transport and forces in confined AOUP systems. In panels (A) and (B), the local diffusion constant reflects the combined effect of persistence time $\tau$, interaction strength $V_0$, and crowding $N$: increasing $\tau$ enhances nonequilibrium persistence, leading to spatially heterogeneous diffusion that deviates from the constant passive value, with stronger suppression near regions of large restoring force. In the harmonic trap (A), this manifests as a reduction of the local diffusion near the trap center for large $V_0$ and $N$, due to caging and interaction-induced hindrance, while for larger $\tau$ the diffusion becomes more position dependent because active correlations couple to the trap curvature. In the double-well potential (B), the effect is more pronounced: diffusion is strongly reduced near the barrier and well minima as $V_0$ and $N$ increase, reflecting crowding-induced trapping and reduced barrier-crossing efficiency, while larger 
$\tau$ enhances accumulation and sharpens spatial contrasts. Panels (C) and (D) show the corresponding effective force, which deviates from the bare conservative force due to activity–interaction coupling: in the harmonic trap (C), increasing 
$\tau$ and $V_0$ renormalizes the force slope, typically softening the effective confinement at moderate activity but leading to nonlinear distortions at high density.

In the double-well case (D), the effective force becomes highly nontrivial, with interaction- and activity-induced modifications that flatten the barrier and skew the force near the wells, signaling an effective nonequilibrium potential shaped jointly by persistence, excluded-volume interactions, and particle number.

\section{Steady-state velocities}
\label{sec:steadystatevelocities}

\begin{figure*}[!tbh]
  \includegraphics[width=\linewidth]{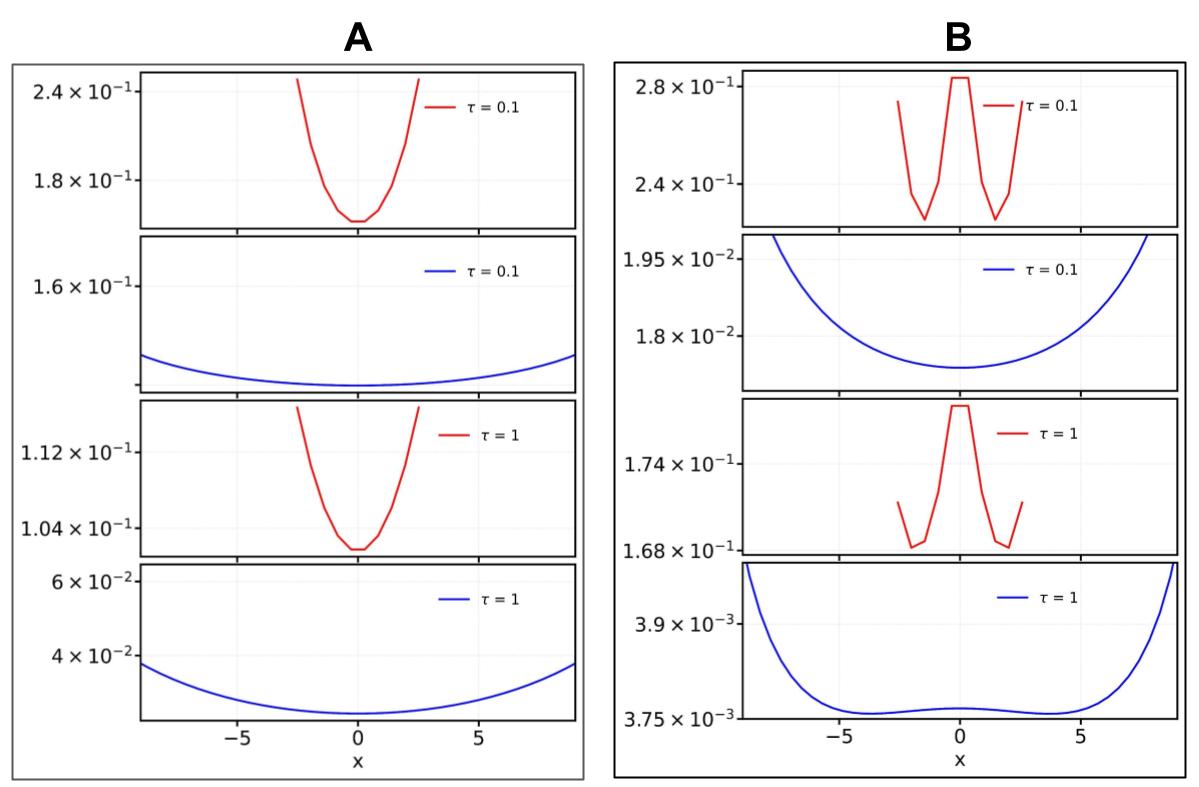}
  \caption{Spatial variation of the steady-state velocity fluctuations in (A) a harmonic trap and (B) a double-well potential. Red curves correspond to a system of $N=10$ particles, while blue curves correspond to $N=50$ particles. Fluctuations are suppressed near the minima of the potentials due to crowding.
}
  \label{fig:velocity_correlations}
\end{figure*}

\begin{figure*}[!tbh]
  \includegraphics[width=\linewidth]{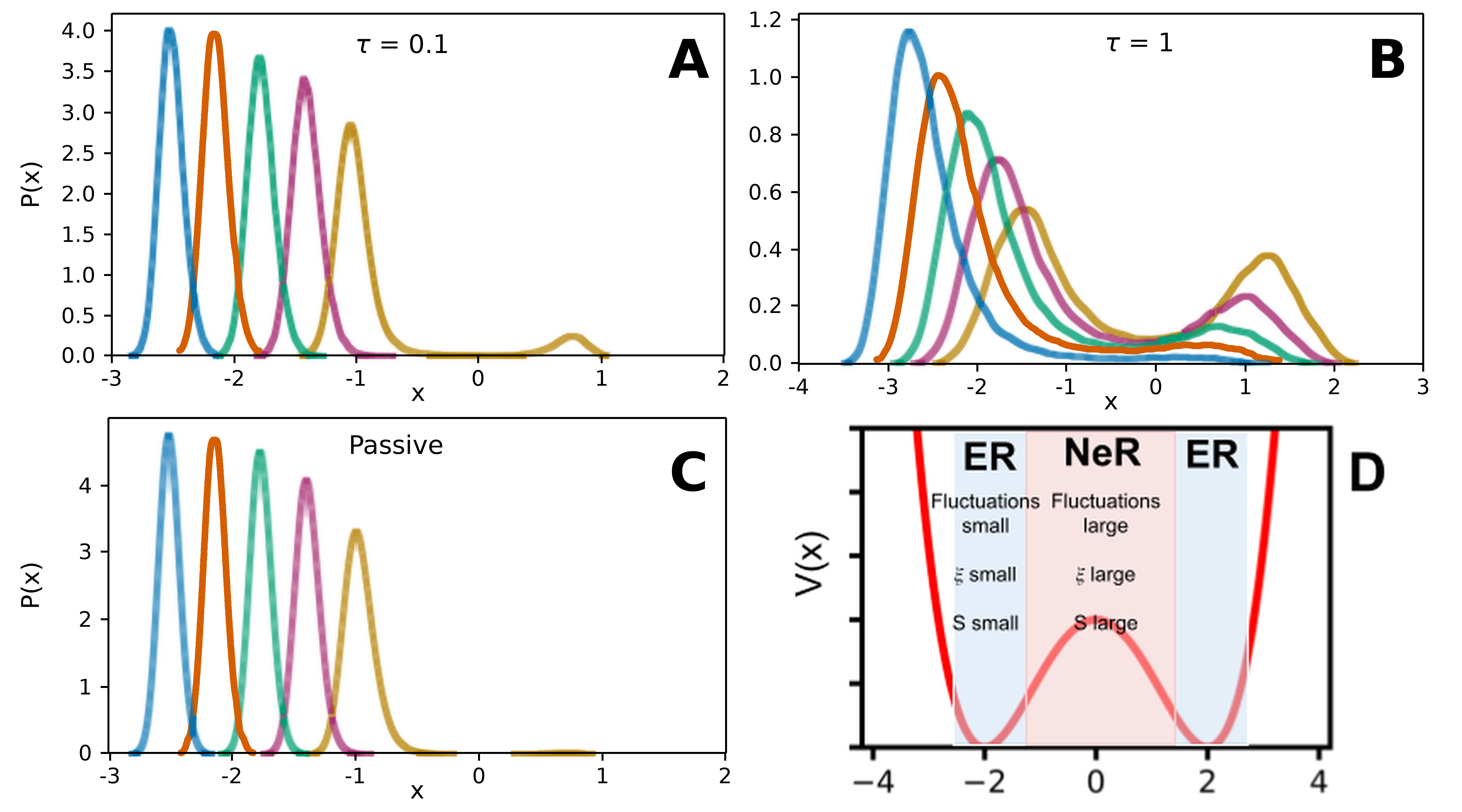}
  \caption{(A-C) Steady-state single-particle position distributions $P_i(x)$ for five interacting AOUPs out of $N = 10$ particles in the left well of a double-well potential. (Adapted from \cite{Parashar2026_aoup}) The distributions are symmetric for the particles in the right well and are not shown. These distributions were obtained from simulations of Eq.\eqref{aoup} with hard potential interactions using the Euler–Maruyama method with a time step $dt = 10^{-3}$ over $10^7$ iterations. Increasing persistence broadens the distributions and induces bimodality for particles located closer to the barrier, reflecting enhanced barrier exploration and nonequilibrium fluctuations. (D) Schematic illustration of the equilibrium regions (ERs) near the well minima and the nonequilibrium region (NeR) near the barrier. ERs are characterized by small fluctuations, short correlation lengths, and low entropy production, whereas the NeR exhibits enhanced fluctuations, long-ranged correlations, and large entropy production due to persistent active motion.
}
  \label{fig:individual_particles}
\end{figure*}

The interplay between activity, confinement, and steric interactions produces a spatially heterogeneous nonequilibrium steady-state. For a single AOUP in a double-well potential, Caprini \textit{et al.}~\cite{caprini2019activejcp} showed that the barrier region is characterized by strong violations of equilibrium behavior. In this region the mobility becomes negative, entropy production is large, and the local velocity distribution develops a bimodal structure. These regions were termed Nonequilibrium Regions (NeR). In contrast, the vicinity of the potential minima behaves approximately as an Equilibrium Region (ER), where the velocity distribution remains nearly Gaussian and entropy production is comparatively small. Here we extend this picture to interacting many-particle systems and investigate how fluctuations and correlations are organized by the local curvature of the confining landscape \rev{, thereby deriving analytically the behaviour we previously obtained from
simulations \cite{Parashar2026_aoup}.}

The particles interact through a Weeks-Chandler-Andersen (WCA) potential, which prevents particle crossings and preserves the ordering of particles throughout the dynamics. As a consequence, each particle index acts as a proxy for its spatial location within the trap. To understand the origin of the observed fluctuations, we linearize Eq.~\eqref{langevin} around the energy-minimized configuration $\{x_i^\ast\}$ and obtain

\begin{align}
\dot{\mathbf{x}} &= \mathbf{v},\\
\dot{\mathbf{v}}
&=
-\left[
\frac{1}{\tau}\mathbf{I}
+\frac{1}{\gamma}\mathbf{H}
\right]\mathbf{v}
-\frac{1}{\tau\gamma}\mathbf{H}\,\delta\mathbf{x}
+\frac{\sqrt{2D}}{\tau}\boldsymbol{\xi},
\label{linearization}
\end{align}

where $\delta x_i=x_i-x_i^\ast$ and

\begin{equation}
H_{ij}
=
\phi''(x_i^\ast)\delta_{ij}
+\partial_{x_i}\partial_{x_j}
V(x_i^\ast-x_j^\ast)
\label{hessianmatrix}
\end{equation}

is the Hessian matrix of the many-body potential energy landscape. The eigenvalues $\lambda_\alpha$ and eigenvectors $\psi_\alpha$ of $\mathbf{H}$ characterize the local stiffness and collective normal modes of the system. In the absence of interactions, $\lambda_\alpha=\phi''(x_\alpha)$ reduces to the local curvature of the external potential. Large positive eigenvalues correspond to stiff, strongly confining regions, whereas small eigenvalues identify soft regions where particles can fluctuate over larger distances. In a double-well potential, the smallest eigenvalues are localized near the barrier, while the largest eigenvalues occur near the well bottoms.

The velocity and position correlation functions are given by

\begin{equation}
\langle v_i(t)v_j(0)\rangle
=
\sum_\alpha
\frac{v_0^2\gamma}
{\gamma+\tau\lambda_\alpha}
\,
\frac{
\frac{\lambda_\alpha}{\gamma}
e^{-\lambda_\alpha t/\gamma}
-
\frac1\tau e^{-t/\tau}
}{
\frac{\lambda_\alpha}{\gamma}
-\frac1\tau
}
\psi_\alpha^i\psi_\alpha^j ,
\label{velocitycorrelations}
\end{equation}

and

\begin{widetext}
\begin{equation}
\langle
\delta x_i(t)\delta x_j(0)
\rangle
=
\sum_\alpha
\frac{
v_0^2\tau\gamma^2
}
{
\lambda_\alpha
(\gamma+\tau\lambda_\alpha)
}
\,
\frac{
\frac{\lambda_\alpha}{\gamma}
e^{-t/\tau}
-
\frac1\tau
e^{-\lambda_\alpha t/\gamma}
}{
\frac{\lambda_\alpha}{\gamma}
-\frac1\tau
}
\psi_\alpha^i\psi_\alpha^j .
\label{positioncorrelations}
\end{equation}
\end{widetext}

\rev{
The functional forms of Eqs.~\eqref{velocitycorrelations} and
\eqref{positioncorrelations} are closely related to the established correlation functions for harmonically confined degrees of freedom
driven by exponentially correlated active noise
\cite{szamel2014self,samanta2016chain}. In particular, Samanta and Chakrabarti extended the single-particle
harmonic calculation to the normal modes of Rouse, Zimm, and semiflexible polymer chains, for which each normal mode obeys an effective linear harmonic equation \cite{samanta2016chain}.
We therefore regard the corresponding harmonic-mode limit of
Eqs.~\eqref{velocitycorrelations}--\eqref{fluctuation} as a consistency check of the present formalism rather than as a new result. The distinction of the present calculation is not simply the passage
from a single degree of freedom to a many-body system. Rather, the
correlations are expressed in terms of the complete spectrum
$\{\lambda_\alpha,\psi_\alpha\}$ of the configuration-dependent
many-body Hessian, Eq.~\eqref{hessianmatrix},
which contains simultaneously the local curvature of the external
confinement and the stiffness generated by interparticle interactions.
Thus, unlike a prescribed set of harmonic normal modes, the collective
mode spectrum in the present system depends on both the particle
configuration and the local geometry of the confining potential. The eigenvalues $\lambda_\alpha$ determine the stiffness and relaxation
of the collective modes, while the corresponding eigenvectors
$\psi_\alpha^i$ describe their spatial structure across the confined system. Equations~\eqref{velocitycorrelations}--\eqref{fluctuation}
therefore allow the fluctuations and correlations to be resolved in
space and make it possible to identify where soft and stiff collective modes are concentrated within the trap. For a harmonic external potential, $\Phi''(x)=k$ is spatially uniform,
so the confinement itself does not generate any spatial variation of
the local curvature. An interacting harmonic system may nevertheless
possess a nontrivial many-body spectrum because of the interaction
contribution to the Hessian. The double-well confinement introduces an
additional and qualitatively distinct source of spatial heterogeneity:
its external curvature varies strongly with position, from positive
curvature near the minima to soft or negative curvature near the
barrier. Comparison of the harmonic and double-well cases therefore
allows us to separate collective effects generated by interactions from
those associated specifically with spatially nonuniform confinement.
}

The corresponding equal-time fluctuations are
\begin{widetext}
\begin{equation}
\langle v_i^2\rangle
=
\sum_\alpha
\frac{v_0^2\gamma}
{\gamma+\tau\lambda_\alpha}
(\psi_\alpha^i)^2,
\qquad
\langle
\delta x_i^2
\rangle
=
\sum_\alpha
\frac{
v_0^2\tau\gamma^2
}
{
\lambda_\alpha
(\gamma+\tau\lambda_\alpha)
}
(\psi_\alpha^i)^2.
\label{fluctuation}
\end{equation}
\end{widetext}

Equation~\eqref{fluctuation} immediately reveals the role of curvature. The contribution of each mode is inversely proportional to $\gamma+\tau\lambda_\alpha$, implying that stiff modes with large positive curvature suppress fluctuations while soft modes enhance them. Therefore, the spatial variation of the fluctuations directly reflects the underlying curvature landscape experienced by the particles.

Fig.~\ref{fig:velocity_correlations} illustrates this behavior for interacting AOUPs confined in harmonic and double-well potentials. In the harmonic trap (Fig.~\ref{fig:velocity_correlations}A), the curvature is positive and approximately uniform throughout the system. Consequently, the velocity fluctuations vary smoothly across space and exhibit a minimum at the trap center. This minimum arises primarily from crowding: particles accumulate near the center of the trap, increasing the effective local stiffness through steric interactions and suppressing fluctuations. Since the curvature does not vary strongly with position, no sharply localized nonequilibrium region emerges.

The double-well potential exhibits qualitatively different behavior (Fig.~\ref{fig:velocity_correlations}B). Near the well bottoms, where the curvature is strongly positive, particles are trapped by large restoring forces and the velocity fluctuations are strongly suppressed. These regions correspond to the ER identified in Ref.~\cite{caprini2019activejcp}. In contrast, the barrier region is characterized by small or negative curvature. The restoring force becomes weak, allowing the persistent active force to dominate the dynamics. Consequently, velocity fluctuations are significantly enhanced and the system enters a NeR. The contrast between ER and NeR becomes particularly evident for small particle numbers ($N=10$). The fluctuation minima coincide with the well bottoms, while a pronounced maximum develops at the barrier. This enhancement reflects the fact that particles approaching the barrier can maintain their active velocities over longer times before being redirected by the external force. For larger systems ($N=50$), increasing particle density stiffens the collective modes, with larger $\lambda_{\alpha}$ suppressing fluctuations throughout the trap. The enhanced effective stiffness generated by neighboring particles reduces the amplitude of active motion and partially masks the curvature-induced heterogeneity. Thus increasing particle number drives the system toward a more equilibrium-like state despite the presence of activity. Increasing $\tau$ strengthens the memory of the active force and reduces the efficiency of local relaxation. From Eq.~\eqref{fluctuation}, modes with small $\lambda_\alpha$ (NeR regions) receive the largest enhancement because the denominator $\gamma+\tau\lambda_\alpha$ grows slowly. As a result, increasing $\tau$ preferentially amplifies fluctuations in the NeR while producing only modest changes in the ER. This selective enhancement of barrier fluctuations leads to increased barrier-crossing events at high persistence.

The differences between the two regions also emerge from the velocity and position correlation functions in Eq.~\eqref{velocitycorrelations} and Eq.\eqref{positioncorrelations}. Two timescales govern the dynamics, $\tau$ and $\frac{\gamma}{\lambda_\alpha}$. The persistence time $\tau$ characterizes the memory of the active force, while $\frac{\gamma}{\lambda_\alpha}$ is the relaxation time associated with a normal mode of stiffness $\lambda_\alpha$. In the ER $\frac{\gamma}{\lambda_\alpha}$ is short and correlations decay rapidly. The dynamics therefore resembles that of an equilibrium system with weak temporal memory. In the NeR, however, $\lambda_\alpha$ becomes small and $\frac{\gamma}{\lambda_\alpha}$ grows substantially. Correlations persist for much longer times, producing coherent motion and enhanced collective fluctuations. The spatial extent of these correlations can be estimated by expanding the Hessian spectrum around a locally homogeneous region, $\lambda(k)
= \phi''(x^\ast)+Kk^2$, where $K$ is determined by the interaction stiffness. Substituting this spectrum into Eq.~\eqref{fluctuation} gives
\begin{equation}
\langle |v(k)|^2\rangle =
\frac{v_0^2\gamma}
{\gamma+\tau\left[\phi''(x^\ast)+Kk^2\right]}.
\end{equation}
Fourier transformation yields an exponentially decaying spatial correlation function with correlation length

\begin{equation}
\xi
=
\sqrt{
\frac{\tau K}
{\gamma+\tau\phi''(x^\ast)}
}.
\label{correlationlength}
\end{equation}
Eq.~\eqref{correlationlength} demonstrates explicitly that the correlation length is controlled by both persistence and curvature. Near the barrier (NeR) as shown in Fig.\ref{fig:individual_particles}-D, where $\phi''(x^\ast)$ is small or negative, $\xi$ becomes large and particles remain correlated over longer distances. Near the well bottoms (ER), where the curvature is strongly positive, the correlation length decreases and collective motion is rapidly suppressed. Increasing $\tau$ enlarges $\xi$, promoting synchronized motion and cooperative barrier-crossing events. $\xi \sim 0$ for passive systems.

\section{Structure factor}
\label{sec:structurefactor}

Now, we analyze how the curvature of the trap and the activity of the particles dictate their density fluctuations in the different regions of the confining potential. The dynamical structure factor quantifies the spectrum of density-density correlations in both space and frequency. In steady-state, the dynamical structure factor of $N$ interacting AOUP particles is derived (see Appendix \ref{Appendix:structure factor} for details) as
\begin{equation}
S(k,\omega) = 
\frac{2Dk^2}{N\tau^2}
\sum_\alpha
\frac{|A_\alpha(k)|^2}
{
\left(
\frac{\lambda_\alpha}{\tau\gamma}-
\omega^2
\right)^2
+
\omega^2
\left(
\frac{1}{\tau}
+
\frac{\lambda_\alpha}{\gamma}
\right)^2
},
\label{dynamic_structure_factor}
\end{equation}
where
\begin{equation}
A_\alpha(k) = 
\sum_i
e^{-ikx_i^\ast}\psi_\alpha(i),
\end{equation}
and ${\lambda_\alpha}$ and ${\psi_\alpha}$ are the eigenvalues and eigenvectors of the Hessian matrix $\mathbf H$ discussed in Sec.\ref{sec:steadystatevelocities}. The eigenvalues $\lambda_{\alpha}$ determine the relaxation frequencies of the collective modes, while the eigenvectors enter through the form factor $A_{\alpha}(k)$, which controls how strongly each mode contributes at a given wavevector $k$. Soft modes with small ${\lambda_\alpha}$, typically associated with particles near the barrier (NeR), generate enhanced low-frequency spectral weight and slow relaxation, reflecting density transitions between the wells and long-lived collective rearrangements. Stiff modes with large ${\lambda_\alpha}$, associated with strongly confined particles near the minima, lead to faster relaxation, similar to equilibrium systems. Increasing $N$ increases the particle density and stiffens the Hessian eigenvalues $\lambda_\alpha$, density fluctuations become increasingly distributed among many modes and the low-frequency response is suppressed leading to more equilibrium-like behavior. In contrast, for small $N$, a few low-lying modes can dominate the sum, leading to stronger fluctuations and more pronounced spectral features. The difference between passive and active systems arises through the persistence time $\tau$. In the passive limit ($\tau\rightarrow 0$), each mode has only the curvature-controlled
relaxation rate $\lambda_\alpha/\gamma$,  whereas the active system contains two timescales: the structural relaxation time $\gamma/\lambda_\alpha$ and the
active persistence time $\tau$. In the limit $\tau \rightarrow 0$, the active
memory relaxes rapidly and the passive Lorentzian form is recovered. For finite
$\tau$, however, soft modes with small $\lambda_\alpha$ are strongly affected by persistence and produce enhanced low-frequency spectral weight. These modes are especially important near the barrier of the double-well potential, whereas near the well minima large $\lambda_\alpha$ suppresses persistent fluctuations and the response becomes closer to the passive one. In harmonic confinement, the curvature is uniform, the Hessian spectrum is relatively regular, and there are no strongly localized barrier modes; therefore active persistence mainly broadens or shifts the relaxation spectrum compared with the passive case, but the response remains smooth and equilibrium-like. \rev{The harmonic case therefore acts here as the spectral baseline: it shows what activity alone does to $S(k,\omega)$ when the curvature does not vary. The additional low-frequency weight seen in the double well, which is absent from this baseline at the same $\tau$, $V_0$ and $N$, is consequently a curvature-induced effect and constitutes a direct spectral signature of the soft barrier modes.}

\section{Entropy production}
\label{sec:entropyproduction}

AOUPs are intrinsically nonequilibrium systems because their persistent self-propulsion continuously injects energy into the dynamics and breaks detailed balance. As a consequence, the system exhibits irreversible probability currents and violates time-reversal symmetry. The steady-state entropy production provides a quantitative measure of this irreversibility and therefore characterizes the degree to which the system departs from equilibrium. In this section, we investigate how confinement generates spatially heterogeneous entropy production and how this heterogeneity is controlled by the local curvature of the confining potential, particle interactions, and persistence time. We used the Spinny-Ford \cite{Spinney2012Entropy} formula to obtain the steady-state entropy production rate in the mean field approximation
\begin{widetext}
\begin{equation}
\dot S_{\mathrm{tot}}
=
\frac{D\tau^2}{4}
\left\langle
\left(
\sum_{i,j}
\partial_{ijj}\varphi
\right)^2
\right\rangle
+
\mathcal O(\tau^3)
\nonumber\
=
\frac{D\tau^2}{4}
\int dx,\rho(x)
\left[
\partial_x^3
\varphi_{\mathrm{eff}}(x)
\right]^2
+
\mathcal O(\tau^3),
\label{entropy}
\end{equation}
\end{widetext}
where the effective potential is
\begin{equation}
\varphi_{\mathrm{eff}}(x) = 
\Phi(x)
+
2\int dx'\rho(x')V(x-x').
\end{equation}
Equation~\eqref{entropy} shows that entropy production is governed by the square of the third derivative of the effective potential. Since $\partial_x^3\varphi_{\mathrm{eff}}$ measures the spatial variation of the local curvature, entropy production is generated only in regions where the curvature changes with position. Thus, entropy production provides a direct link between the nonequilibrium dynamics and the geometry of the effective energy landscape.

The leading-order dependence on persistence follows directly from Eq.~\eqref{entropy},
\begin{equation}
\dot S_{\mathrm{tot}}\propto D\tau^2.
\end{equation}
Consequently, the passive limit $\tau\rightarrow0$ restores equilibrium behavior and causes the entropy production to vanish. In this limit, the active force loses its temporal memory and reduces to thermal white noise, recovering detailed balance and time-reversal symmetry.

The role of confinement becomes particularly evident when comparing harmonic and double-well potentials. \rev{This comparison is the sharpest illustration of why the harmonic trap is retained as a reference system: Eq.~\eqref{entropy} depends on the confinement only through $\partial_x^3\varphi_{\rm eff}$, so a constant-curvature trap is, at this order, the exactly solvable null case in which activity is present but produces no irreversibility whatsoever. Any entropy production observed in the double well at the same activity is therefore unambiguously attributable to curvature variation, and any entropy production observed in the interacting harmonic trap is unambiguously attributable to the interactions.} For a harmonic trap, the curvature is uniform, $\Phi''(x)=k$ and therefore $\Phi'''(x)=0$. In the absence of interactions, the entropy production vanishes to order $\tau^2$, despite the presence of active forces. The dynamics therefore remain effectively equilibrium-like. Interparticle interactions introduce a weak spatial dependence through the density-dependent contribution to $\varphi_{\mathrm{eff}}(x)$, producing a small but finite entropy production, primarily near the edges of the particle cloud where the density profile varies most strongly.

In a double-well potential, Eq.~\eqref{entropy_particle} demonstrates that the entropy production of a given particle depends on how strongly its position distribution overlaps with the nonequilibrium regions of the trap. We simulate the single particle distribution $\rho_i(x) = 
\langle\delta(x_i(t)-x)\rangle$ in Fig.\ref{fig:individual_particles}. For the i$^\text{th}$ particle the entropy production rate is
\begin{equation}
\dot S_i = 
\frac{D\tau^2}{4}
\int dx\rho_i(x)
\left[
\partial_x^3\varphi_{\mathrm{eff}}(x)
\right]^2
+
\mathcal O(\tau^3).
\label{entropy_particle}
\end{equation}
The density in Eq.\eqref{entropy} is related to the single particle distribution by $\rho(x) = 
\sum_i\rho_i(x)$. Since $\partial_x^3\varphi_{\text{eff}}(x) = 6kx$, from Eq.\eqref{entropy_particle} we obtain $\dot S_i = \frac{D\tau^2}{4}\langle x^2\rangle$ i.e. depends on the width of the single particle distribution $\rho_i$ or the variance of the particle position.  The entropy production is therefore directly related to the spatial extent of the particle distribution. Particles localized near the minima of the wells have narrow, unimodal distributions with small positional variance and consequently generate little entropy. These particles predominantly occupy the equilibrium-like regions (ERs), where the local curvature varies slowly and fluctuations remain weak. In contrast, particles near the barrier  exhibit broader and often bimodal distributions, reflecting higher entropy production.

In Fig.~\ref{fig:individual_particles}, for small persistence times, the distributions remain narrow and localized near the minima, leading to weak entropy production. As $\tau$ increases, the distributions broaden and develop pronounced bimodality, indicating that particles increasingly explore both wells and the barrier region separating them. This broadening increases the positional variance $\langle x^2\rangle_i$ and consequently enhances the entropy production. Therefore, the increase in entropy production with persistence arises from two complementary mechanisms: the explicit $D\tau^2$ dependence in Eq.~\eqref{entropy} and the increased sampling of the nonequilibrium barrier region caused by broader, more strongly bimodal particle distributions.

\section{Conclusion}
\label{sec:conclusion}

In this work, we investigated the steady-state dynamics, density fluctuations, velocity correlations, structure factor, and entropy production of interacting active Ornstein-Uhlenbeck particles confined in harmonic and double-well potentials. The central novelty of the study is the identification of how activity, interactions, particle number, and the local curvature of the confining potential together organize the system into equilibrium-like and nonequilibrium-like spatial regions. \rev{
The novelty of the present work does not lie in the correlation functions
of harmonically confined degrees of freedom driven by exponentially
correlated active noise, which are well established
\cite{szamel2014self,samanta2016chain}. In particular, the latter work
also extends the harmonic active-noise description to the normal modes of
polymer chains. Rather, the present contribution is the analytical
description, within the stated mean-field, small-persistence, and
linearized approximations, of an interacting confined AOUP system in
which steric interactions, persistent active forcing, and spatially
varying external curvature act simultaneously. For the interacting system, the velocity and position correlations,
Eqs.~\eqref{velocitycorrelations}--\eqref{fluctuation}, are resolved in
terms of the eigenvalues and eigenvectors of the configuration-dependent
many-body Hessian. Its spectrum contains contributions from both the
local curvature of the external confinement and the interparticle
interactions, rather than from prescribed harmonic mode stiffnesses
alone. Within a locally homogeneous description, this framework further
leads to the curvature-dependent correlation length,
Eq.~\eqref{correlationlength}, and to the dynamical structure factor,
Eq.~\eqref{dynamic_structure_factor}. In addition, the small-persistence
expansion yields the leading-order curvature-gradient contribution to the
entropy production, Eq.~\eqref{entropy}, thereby relating irreversibility
to spatial variations of the effective curvature. The harmonic trap is therefore used as a constant-external-curvature
reference system rather than as a new active-matter problem. Although
interactions can still generate a nontrivial many-body mode spectrum in
this case, the external curvature remains spatially uniform. The
double-well potential provides the complementary situation in which the
external curvature changes strongly between the well minima and the
barrier region. Comparing these two confinements consequently allows us
to separate interaction-induced collective effects from those associated
specifically with spatially nonuniform confinement and soft barrier
modes. In this way, the present analysis identifies how persistence,
crowding, and local confinement geometry jointly control density
fluctuations, collective correlations, mode softening, and
irreversibility in interacting confined active systems.
} While the harmonic trap remains largely equilibrium-like due to its uniform positive curvature, the double-well potential produces strong spatial heterogeneity, with the well minima behaving as equilibrium regions and the barrier acting as a nonequilibrium region.

Starting from the many-particle AOUP dynamics, we derived a mean-field single-particle Fokker-Planck equation and obtained an effective density equation with position-dependent diffusion and effective force. This formulation shows that active persistence and excluded-volume interactions renormalize both the confining potential and the local transport properties. In the harmonic trap, activity broadens the density profile and interactions suppress diffusion in crowded regions. In the double-well potential, persistence reshapes the steady-state density by enhancing barrier exploration and modifying the population contrast between the wells and the barrier.

A major finding is that the nonequilibrium character of the system is controlled by the local curvature landscape. Linearizing the dynamics around the minimum-energy configuration shows that the eigenvalues of the Hessian matrix determine the amplitude and lifetime of velocity and position fluctuations. Stiff modes localized near the well minima suppress fluctuations and produce equilibrium-like behavior, whereas soft modes near the barrier enhance fluctuations, correlations, and collective motion. This provides a direct microscopic explanation for the emergence of nonequilibrium regions in the double-well trap. Increasing the persistence time strengthens these soft-mode fluctuations and increases the spatial correlation length, thereby promoting cooperative and synchronized barrier-crossing events. In contrast, increasing the particle number enhances crowding and collective stiffness, suppressing fluctuations and driving the system toward a more equilibrium-like state.

The dynamical structure factor further reveals how active persistence modifies density fluctuations. Soft barrier modes generate enhanced low-frequency spectral weight and slow relaxation, while stiff modes near the well bottoms relax rapidly and resemble passive equilibrium modes. Thus, the structure factor provides a spectral signature of the spatial separation between equilibrium-like and nonequilibrium-like regions.

Finally, we showed that entropy production is governed by the spatial variation of the effective curvature, through the third derivative of the effective potential. Consequently, entropy production vanishes to leading order in a noninteracting harmonic trap but becomes finite in the double-well potential and in interacting systems where the effective landscape is spatially inhomogeneous. Particles localized near the barrier region exhibit  higher entropy production, whereas those near the well minima contribute weakly.

Overall, our results demonstrate that confined active matter cannot be characterized only by global activity parameters. Instead, nonequilibrium behavior is highly localized and is selected by the geometry of the effective energy landscape. The interplay of persistence, curvature, and crowding interactions produces spatially resolved transport heterogeneity, enhanced barrier fluctuations, long-ranged correlations, and localized entropy production. These findings provide a unified framework for understanding how confinement controls collective nonequilibrium dynamics in interacting active systems.

\section*{Acknowledgements}
Author SD  acknowledges financial support by Science
and Engineering Research Board (SERB), Government of
India (File SRG/2022/000598) and (File MTR/2022/000281).
The author also acknowledges financial support from the
Birla Institute of Technology and Science, Pilani (File
PLN/AD/2022-23/3). The author also thanks DST-FIST for
the computational resources provided to the Department of
Physics, BITS Pilani.

\section*{Declaration}
During the preparation of this manuscript, the authors used ChatGPT and Grammarly for the sole purpose of English language polishing and grammar correction. After using this tool, the authors reviewed and edited the content as needed and take full responsibility for the content of the publication.

\appendix
\section{Derivation of Eq.\eqref{bbgkymeanfield}}
\label{Appendix:Fokker-Planck}
For a general stochastic equation of motion of the form
\begin{equation}
\dot{y} = A_i(y,t) + \sum_kB_{ik}(y,t)\epsilon_k,
\end{equation}
where $\epsilon$'s satisfy
\begin{equation}
\langle\epsilon_k\rangle = 0, \langle\epsilon_k\epsilon_l\rangle = \delta_{kl}\delta(t_1-t_2),
\end{equation}
the Fokker Planck equation is given by
\begin{widetext}
\begin{equation}
\dot{P} = -\sum_i\partial_{y_i}\biggl\{\biggl[A_i+\frac{1}{2}\sum_{jk}B_{jk}\partial_{y_i}B_{ik}\biggr]P\biggr\}+\frac{1}{2}\sum_{ij}\partial_{y_i}\partial_{y_j}\biggl\{\biggl[\sum_{k}B_{ik}B_{jk}\biggr]P\biggr\}.
\end{equation}
\end{widetext}
For our system in Eq.\eqref{langevin} we get
\begin{equation}
\partial_t P
=-\sum_i \partial_{x_i}\!\left(\nu_i P\right) 
-\sum_i \partial_{\nu_i}\left[A_iP\right] 
+\frac{D}{\tau^2}\sum_i \partial_{\nu_i}^2 P,
\label{fokkerplanck}
\end{equation}
where 
\begin{widetext}
\begin{equation}
A_i(\mathbf{x},\boldsymbol{\nu}) =
-\sum_j\Bigg[
\frac{1}{\tau}\delta_{ij}
+\frac{1}{\gamma}\Phi''(x_i)\delta_{ij}
+\frac{1}{\gamma}\partial_{x_i x_j}V(x_i-x_j)
\Bigg]\nu_j 
-\frac{1}{\gamma\tau}\Phi'(x_i)
+\frac{1}{\gamma\tau}F_i.
\label{firction}
\end{equation}
\end{widetext}

The first equation of the BBGKY hierarchy for the reduced distribution functions is
\begin{widetext}
\begin{align}
\dot{P}_1(x,\nu) &  = -\partial_x[\nu P_1(x,\nu)]+\partial_{\nu}\biggl[\frac{1}{\tau}\nu P_1(x,\nu)\biggr]+\partial_{\nu}\biggl[\int dx^{\prime}d\nu^{\prime}\frac{1}{\gamma}\partial_{xx^{\prime}}V(x-x^{\prime})\nu^{\prime}P_2(x,\nu,x^{\prime},\nu^{\prime})\biggr] \nonumber \\ & +\partial_{\nu}\biggl[\frac{1}{\gamma}\nu \partial_{x}^2\Phi(x)P_1(x,\nu)\biggr] +\frac{1}{\gamma\tau}{\partial_x\Phi(x)}\partial_{\nu} P_1(x,\nu)+\frac{1}{\gamma\tau}\int dx^{\prime}d\nu^{\prime}{\partial_xV(x-x^{\prime})}\nonumber\\&\partial_{\nu}P_2(x,\nu,x^{\prime},\nu^{\prime})+\frac{D}{\tau^2}\partial_{\nu}^2P_1(x,\nu).
\label{bbgky}
\end{align}
\end{widetext}
$P_1(x,\nu)$ is the joint distribution of the position $x$ and the "velocity" $\nu$. $P_2(x,\nu, x^{\prime},\nu^{\prime})$ is the two-body correlation function.

\rev{Equation \eqref{bbgky} is obtained from Eq.~\eqref{fokkerplanck} by integrating over the positions and velocities of the $N-1$ particles $i=2,\dots,N$, using the reduced distributions
\begin{widetext}
\begin{align}
P_1(x,\nu) & = \int\prod_{i=2}^{N}dx_id\nu_i\, P(x,x_2,\dots;\nu,\nu_2,\dots),\nonumber\\
P_2(x,\nu,x^{\prime},\nu^{\prime}) & = \int\prod_{i=3}^{N}dx_id\nu_i\, P(x,x^{\prime},x_3,\dots;\nu,\nu^{\prime},\nu_3,\dots).
\label{reduceddistributions}
\end{align}
\end{widetext}
Every term of Eq.~\eqref{fokkerplanck} that is a total derivative with respect to an integrated variable contributes only a boundary term and are neglected. Eq.~\eqref{bbgky} is not closed as because $\dot P_1$ contains $P_2$.}
In order to solve for $P_2$ we need to consider the next equation of the hierarchy. However here we use the mean field approximation $P_2(x,\nu, x^{\prime},\nu^{\prime}) = P_1(x,\nu)P_1( x^{\prime},\nu^{\prime})$ which closes Eq.\eqref{bbgky} and we get Eq.\eqref{bbgkymeanfield}. \rev{Substituting this factorisation into the two $P_2$ terms of Eq.~\eqref{bbgky} replaces them by the mean interaction field generated by the one-body density, $\int dx^{\prime}d\nu^{\prime}\,\partial_{xx^{\prime}}V(x-x^{\prime})\nu^{\prime}P_1(x,\nu)P_1(x^{\prime},\nu^{\prime})$ and $\int dx^{\prime}d\nu^{\prime}\,\partial_xV(x-x^{\prime})\partial_\nu P_1(x,\nu)P_1(x^{\prime},\nu^{\prime})$ respectively, which is precisely the mean-field (two-body factorisation) closure discussed in Sec.~\ref{sec:aoup}; its conditions of validity are stated there.}

To extract macroscopic observables, we compute moments of 
Eq.\eqref{bbgkymeanfield} over $\nu$. We obtain the continuity equation for the particle density $\rho(x, t)=\int d\nu P_1(x,\nu, t)$
\begin{equation}
\partial_t\rho(x, t)  = \partial_xj(x, t),
\label{continuity}
\end{equation}
where $j(x, t) = \int d\nu \nu P_1(x,\nu, t)$ is the current. Similarly, multiplying Eq.\eqref{bbgkymeanfield} by $\nu$ and $\nu^2$ and integrating over $\nu$ we get the following equations for $j(x, t)$ and $\mathcal{V}(x, t) = \int d\nu \nu^2P_1(x,\nu, t)$
\begin{widetext}
\begin{align}
\partial_t j(x, t) & = -\partial_x\mathcal{V}(x, t)-\frac{1}{\tau}j(x, t)-\frac{1}{\gamma}\Phi^{\prime\prime}(x)j(x, t)-\frac{\rho(x, t)}{\gamma}\int dx^{\prime}j(x^{\prime}, t)\partial_{xx^{\prime}}V(x-x^{\prime})-\frac{1}{\gamma\tau}\times\nonumber\\
&\Phi^{\prime}(x)\rho(x, t)-\frac{\rho(x, t)}{\gamma\tau}\int dx^{\prime}{\partial_xV(x-x^{\prime})}\rho(x^{\prime}, t),\label{current}\\
\partial_t\mathcal{V}(x, t) & = -\partial_x\chi(x, t)-\frac{2}{\tau}\mathcal{V}(x, t)-\frac{2}{\gamma}\Phi^{\prime\prime}(x)\mathcal{V}(x, t)-2\frac{j(x, t)}{\gamma}\int dx^{\prime}j(x^{\prime}, t)\partial_{xx^{\prime}}V(x-x^{\prime})\nonumber\\&-\frac{2}{\gamma\tau}\partial_x\Phi(x)j(x, t)-\frac{2}{\gamma\tau}j(x, t)\int dx^{\prime}\partial_xV(x-x^{\prime})\rho(x^{\prime}, t)+\frac{2D}{\tau^2}\rho(x, t),
\label{susceptibility}
\end{align}
\end{widetext}
where $\chi(x, t) = \int d\nu \nu^3P_1(x,\nu, t)$. 

\rev{Equations \eqref{continuity}-\eqref{susceptibility} are the first three members of a velocity-moment hierarchy: the equation for $\rho$ involves $j$, the equation for $j$ involves $\mathcal V$, and the equation for $\mathcal V$ involves $\chi$. The hierarchy is truncated at this order by two assumptions, each of which we now state together with its regime of validity. (i) On time scales long compared with the velocity relaxation time, the second moment is quasi-stationary, $\partial_t\mathcal V(x,t)\simeq 0$; this is the adiabatic (local-equilibrium) elimination of the velocity degrees of freedom and requires $\tau$ and $\gamma/\lambda$ to be short compared with the time scale over which the density evolves. (ii) The third moment is assumed to be slowly varying in space, $\partial_x\chi(x)\simeq 0$, which holds when the velocity distribution is locally close to Gaussian, so that $\chi$ is determined by $\mathcal V$ and inherits its smoothness. Under (i) and (ii), Eq.~\eqref{susceptibility} becomes algebraic in $\mathcal V$ and can be solved for it.}

On long time scales, $\mathcal{V}(x,t)$ has a weak variation with time. In addition, we assume a small gradient of $\chi(x)$ and neglect the $\partial_x\chi(x)$ term. Eq.\eqref{susceptibility} leads to
\begin{widetext}
\begin{align}
\mathcal{V}(x) & =\frac{D}{\tau}\frac{\rho(x)}{1+\frac{\tau}{\gamma}\Phi^{\prime\prime}(x)}-j(x)\Psi(x), \text{ where}\\
\Psi(x) & = \frac{1}{1+\frac{\tau}{\gamma}\Phi^{\prime\prime}(x)}\left[\frac{1}{\gamma}\int dx^{\prime}j(x^{\prime})\partial_{xx^{\prime}}V(x-x^{\prime})+\frac{1}{\gamma}\Phi^{\prime}(x)+\frac{1}{\gamma}\int dx^{\prime}\rho(x^{\prime})\partial_{x}V(x-x^{\prime})\right].\nonumber
\end{align}
\end{widetext}
Similarly the local current $j(x)$
\begin{widetext}
\begin{align}
j(x)&= \frac{1}{\Gamma(x)}\left[D\partial_x\left[\frac{\rho(x)}{1+\frac{\tau}{\gamma}\Phi^{\prime\prime}(x)}\right]+\frac{\rho(x)}{\gamma}\left[\Phi^{\prime}(x)+\int V^{\prime}(x-x^{\prime})\rho(x^{\prime})\right]\right],
\label{appendix:current}
\end{align}
\end{widetext}
where $\Gamma(x) = 1+ \frac{\tau}{\gamma}\Phi^{\prime\prime}(x)-\tau\Psi^{\prime}(x)$. In the derivation we have neglected the gradient of the local current $\partial_xj(x)$. \rev{This last step is the third and final approximation of the derivation: substituting $\mathcal V(x)$ from Eq.~\eqref{susceptibility} into Eq.~\eqref{current} and setting $\partial_t j\simeq 0$ produces a term $\partial_x j$, which is negligible in and near the steady state where the current is small and slowly varying, and which vanishes identically in the zero-current solution $j=0$ used in Sec.~\ref{sec:steadystatedensities}. Comparing Eq.~\eqref{appendix:current} with the constitutive form $j = -\left[D_{\rm eff}(x)\partial_x\rho+\mathcal F(x)\rho\right]$ and inserting the result into the continuity equation \eqref{continuity} then gives Eq.~\eqref{density}, and identifies $D_{\rm eff}(x)$ and $\mathcal F(x)$ as quoted in Eqs.~\eqref{deff}. The positivity condition that $\Gamma(x)$ and $1+\frac{\tau}{\gamma}\Phi^{\prime\prime}(x)$ must satisfy for this identification to be meaningful is discussed in Sec.~\ref{sec:aoup}.}

\section{Correlation functions}
\label{Appendix:correlation functions}

Let the Hessian $\mathbf{H}$ in Eq.\eqref{linearization} be diagonalized as
\begin{equation}
\mathbf{H}\boldsymbol{\psi}^{\alpha}=\lambda_\alpha \boldsymbol{\psi}^{\alpha} .
\end{equation}
Defining the normal modes
\[
\delta x_i(t) = \sum_{\alpha}q_{\alpha}(t)\psi_{\alpha}^i,
\qquad 
v_i(t) = \sum_{\alpha}u_{\alpha}(t)\psi_{\alpha}^i,
\]
the equations of motion in Eq.\eqref{linearization} become
\begin{align}
\dot q_\alpha &= u_\alpha, \\
\dot u_\alpha
&=
-\left(\frac1\tau+\frac{\lambda_\alpha}{\gamma}\right)u_\alpha
-\frac{\lambda_\alpha}{\tau\gamma}q_\alpha
+\frac{\sqrt{2D}}{\tau}\xi_\alpha .
\end{align}

Then
\begin{equation}
\ddot q_\alpha
+(\frac{1}{\tau}+\frac{\lambda_\alpha}{\gamma})\dot q_\alpha
+\frac{\lambda_\alpha}{\tau\gamma} q_\alpha
=
\frac{\sqrt{2D}}{\tau}\xi_\alpha .
\label{modeequation}
\end{equation}

\rev{Equation \eqref{modeequation} is a damped harmonic oscillator driven by white noise. Its homogeneous solutions are $e^{zt}$ with
\begin{widetext}
\begin{equation}
z^2+\left(\frac{1}{\tau}+\frac{\lambda_\alpha}{\gamma}\right)z+\frac{\lambda_\alpha}{\tau\gamma}=0
\quad\Longrightarrow\quad
z_1=-\frac{1}{\tau},\qquad z_2=-\frac{\lambda_\alpha}{\gamma},
\label{characteristicroots}
\end{equation}
\end{widetext}
so that the two relaxation rates of a normal mode are exactly the active persistence rate $1/\tau$ and the structural rate $\lambda_\alpha/\gamma$. This factorisation of the characteristic polynomial is the origin of the two exponentials, $e^{-t/\tau}$ and $e^{-\lambda_\alpha t/\gamma}$, and of the denominator $\frac{\lambda_\alpha}{\gamma}-\frac{1}{\tau}$ appearing in Eqs.~\eqref{velcorrapp} and \eqref{poscorrapp} below; the degenerate case $\lambda_\alpha=\gamma/\tau$ is regular and gives the usual critically damped form $\propto t\,e^{-t/\tau}$.}

\rev{The equal-time moments follow from stationarity. Since $\frac{d}{dt}\langle q_\alpha^2\rangle = 2\langle q_\alpha u_\alpha\rangle$ and $\langle q_\alpha^2\rangle$ is time independent in the steady state, one has $\langle q_\alpha u_\alpha\rangle = 0$. Multiplying Eq.~\eqref{modeequation} by $q_\alpha$ and averaging, the noise term drops out by causality and $\langle q_\alpha\ddot q_\alpha\rangle = -\langle u_\alpha^2\rangle$, which gives $\lambda_\alpha\langle q_\alpha^2\rangle = \gamma\langle u_\alpha^2\rangle$; multiplying instead by $u_\alpha$ and averaging fixes the velocity variance through the noise strength, $\langle u_\alpha^2\rangle = \frac{D\gamma}{\tau(\gamma+\tau\lambda_\alpha)} = \frac{v_0^2\gamma}{\gamma+\tau\lambda_\alpha}$ using $D=v_0^2\tau$. Together these give Eq.~\eqref{equaltimevariance} below without solving the dynamics. The two-time correlators are then obtained from the Onsager regression theorem: the correlation functions decay with the same two rates \eqref{characteristicroots} as the deterministic relaxation, with amplitudes fixed by the equal-time moments just computed, i.e.\ $C^{(v)}_\alpha(0)=\langle u_\alpha^2\rangle$, $\dot C^{(x)}_\alpha(0)=\langle q_\alpha u_\alpha\rangle=0$. Imposing these two conditions on the general combination $c_1e^{-t/\tau}+c_2e^{-\lambda_\alpha t/\gamma}$ determines $c_1$ and $c_2$ uniquely and yields Eqs.~\eqref{velcorrapp} and \eqref{poscorrapp}.}

Using
$\langle q_\alpha u_\alpha\rangle=0$ and $D = v_0^2\tau$, 
we obtain the velocity-velocity correlations as 
\begin{equation}
C^{(v)}_{\alpha}(t) = \langle u_\alpha(t)u_\alpha(0)\rangle
=
\frac{v_0^2\gamma}{\gamma+\tau\lambda_\alpha}
\frac{
\frac{\lambda_\alpha}{\gamma}e^{-\lambda_\alpha t/\gamma}
-
\frac1\tau e^{-t/\tau}
}{
\frac{\lambda_\alpha}{\gamma}-\frac1\tau
}.
\label{velcorrapp}
\end{equation}
Similarly, the position-position correlation is
\begin{widetext}
\begin{equation}
C^{(x)}_{\alpha}(t) =\langle q_\alpha(t)q_\alpha(0)\rangle
=
\frac{v_0^2\tau\gamma^2}
{\lambda_\alpha(\gamma+\tau\lambda_\alpha)}
\frac{
\frac{\lambda_\alpha}{\gamma}e^{-t/\tau}
-
\frac1\tau e^{-\lambda_\alpha t/\gamma}
}{
\frac{\lambda_\alpha}{\gamma}-\frac1\tau
}.
\label{poscorrapp}
\end{equation}
\end{widetext}

From these equations we obtain the equal time variance
\begin{widetext}
\begin{equation}
C^{(v)}_{\alpha}(0) =\langle u_\alpha^2\rangle
=
\frac{v_0^2\gamma}{\gamma+\tau\lambda_\alpha}, \qquad
C^{(x)}_{\alpha}(0) = \langle q_\alpha^2\rangle
=
\frac{D\gamma^2}{\lambda_\alpha(\gamma+\tau\lambda_\alpha)}
=
\frac{v_0^2\tau\gamma^2}
{\lambda_\alpha(\gamma+\tau\lambda_\alpha)}
\label{equaltimevariance}
\end{equation}
\end{widetext}

Therefore,
\begin{equation}
\langle v_i(t)v_j(0)\rangle = \sum_\alpha C^{(v)}_\alpha(t)\psi_\alpha^i \psi_\alpha^j,
\end{equation}
and
\begin{equation}
\langle \delta x_i(t)\delta x_j(0)\rangle
=\sum_\alpha C^{(x)}_\alpha(t)\psi_\alpha^i \psi_\alpha^j.
\end{equation}

\section{Derivation of Structure factor}
\label{Appendix:structure factor}

The microscopic density field is
\begin{equation}
\rho(x,t) = \sum_i \delta(x-x_i(t)),
\end{equation}
with Fourier probe
\begin{equation}
\rho_k(t) = \sum_i e^{-ik x_i(t)} .
\end{equation}
Expanding around the steady-state positions $\{x_i^{\ast}\}$
\begin{equation}
x_i(t) = x_i^* + \delta x_i(t),
\end{equation}
\rev{and expanding the exponential to first order in the displacement,}
\begin{equation}
\rho_k(t)
\simeq
\sum_i e^{-ikx_i^*}
\left[1 - ik\,\delta x_i(t)\right].
\label{linearisedrho}
\end{equation}
\rev{The linearisation in Eq.~\eqref{linearisedrho} is controlled provided the displacements are small on the scale of the probed wavelength,
\begin{equation}
k\sqrt{\langle \delta x_i^2\rangle}\ll 1 ,
\label{linearisationcondition}
\end{equation}
with $\langle\delta x_i^2\rangle$ given by Eq.~\eqref{fluctuation}; the results below therefore describe the small-$k$ (collective) part of the spectrum, and are least accurate for particles in the soft barrier region, where $\langle\delta x_i^2\rangle$ is largest. The $k$-independent first term of Eq.~\eqref{linearisedrho} is the static lattice factor: being time independent, it contributes only an elastic $\delta(\omega)$ contribution to the structure factor and is excluded from the fluctuating part below, which is what $S(k,\omega)$ at $\omega\neq 0$ measures.}

The fluctuating part is therefore
\begin{equation}
\delta \rho_k(t)
=
-ik \sum_i e^{-ikx_i^*}\,\delta x_i(t).
\label{deltarhok}
\end{equation}
Writing the displacement field in terms of the eigenvalues $\lambda_{\alpha}$ and eigenvectors $\psi_{\alpha}$ of the Hessian matrix $\mathbf{H}$ (Appendix \ref{Appendix:correlation functions}) as
\begin{equation}
\delta x_i(t) = \sum_\alpha q_\alpha(t)\,\psi_\alpha(i).
\end{equation}
Each normal mode obeys
\begin{equation}
\ddot{q}_\alpha
+
\left(
\frac{1}{\tau} + \frac{\lambda_\alpha}{\gamma}
\right)\dot{q}_\alpha
+
\frac{\lambda_\alpha}{\tau\gamma}\, q_\alpha
=
\frac{\sqrt{2D}}{\tau}\,\eta_\alpha(t).
\label{modeequationC}
\end{equation}
The dynamical structure factor is defined as
\begin{equation}
S(k,\omega)
=
\frac{1}{N}
\int_{-\infty}^{\infty} dt\,
e^{i\omega t}
\left\langle
\delta\rho_k(t)\,\delta\rho_{-k}(0)
\right\rangle .
\end{equation}

Define the projection of density fluctuations onto mode $\alpha$:
\begin{equation}
A_\alpha(k)
=
\sum_i e^{-ikx_i^*}\,\psi_\alpha(i).
\end{equation}
\rev{Substituting Eq.~\eqref{deltarhok} into the definition of $S(k,\omega)$ gives a double sum over modes, $\sum_{\alpha\beta}A_\alpha(k)A^{\ast}_\beta(k)\langle q_\alpha(\omega)q^{\ast}_\beta(\omega)\rangle$. The normal modes are driven by the independent noises $\eta_\alpha = \sum_i\psi^i_\alpha\xi_i$, which satisfy $\langle\eta_\alpha\eta_\beta\rangle\propto\delta_{\alpha\beta}$ because the eigenvectors of the symmetric Hessian are orthonormal; the modes are therefore statistically independent, $\langle q_\alpha(\omega)q^{\ast}_\beta(\omega)\rangle = \delta_{\alpha\beta}\langle|q_\alpha(\omega)|^2\rangle$, and the double sum collapses to the single mode-diagonal sum}
\begin{equation}
S(k,\omega)
=
\frac{k^2}{N}
\sum_\alpha
|A_\alpha(k)|^2
\left\langle |q_\alpha(\omega)|^2 \right\rangle .
\label{Smodesum}
\end{equation}
\rev{The spectral density of a single mode follows by Fourier transforming Eq.~\eqref{modeequationC}. With $q_\alpha(t)=\int\frac{d\omega}{2\pi}e^{-i\omega t}q_\alpha(\omega)$ the equation becomes algebraic,
\begin{equation}
\left[-\omega^2-i\omega\left(\frac{1}{\tau}+\frac{\lambda_\alpha}{\gamma}\right)+\frac{\lambda_\alpha}{\tau\gamma}\right]q_\alpha(\omega)=\frac{\sqrt{2D}}{\tau}\eta_\alpha(\omega),
\label{fourieredmode}
\end{equation}
so that, using $\langle|\eta_\alpha(\omega)|^2\rangle = 1$ for unit-intensity white noise and taking the squared modulus of the bracket, whose real and imaginary parts are $\frac{\lambda_\alpha}{\tau\gamma}-\omega^2$ and $-\omega(\frac{1}{\tau}+\frac{\lambda_\alpha}{\gamma})$ respectively, one reads off Eq.~\eqref{modespectrum} below.}
Using the following expression for each normal mode
\begin{equation}
\left\langle |q_\alpha(\omega)|^2 \right\rangle
=
\frac{2D/\tau^2}
{
\left(
\frac{\lambda_\alpha}{\tau\gamma} - \omega^2
\right)^2
+
\omega^2
\left(
\frac{1}{\tau} + \frac{\lambda_\alpha}{\gamma}
\right)^2
},
\label{modespectrum}
\end{equation}
we obtain Eq.\eqref{dynamic_structure_factor} in the main text.

\section{Derivation of Entropy production}
\label{Appendix:entropy production}

The probability currents of Eq. \eqref{fokkerplanck} are
\begin{align}
J_{x_i} &= \nu_i P ,\\
J_{\nu_i}
&=
A_i P
-\frac{D}{\tau^2}\partial_{\nu_i}P.
\end{align}
We rewrite  
\begin{equation}
A_i(\mathbf x,\boldsymbol\nu)
=
-\sum_j \Lambda_{ij}(\mathbf x)\nu_j
-\frac{1}{\gamma\tau}\Phi'(x_i)
+\frac{1}{\gamma\tau}F_i,
\end{equation}
with $\Lambda_{ij}(\mathbf x) = \frac{1}{\tau}\delta_{ij}
+\frac{1}{\gamma}\Phi''(x_i)\delta_{ij}
+\frac{1}{\gamma}\partial_{x_i x_j}V(x_i-x_j)$.
Following \cite{Spinney2012Entropy}, under time reversal, we split $A_i$ into reversible and irreversible components
\begin{align}
A_{\nu_i}^{\mathrm{irr}}
&=
-\sum_j \Lambda_{ij}(\mathbf x)\nu_j ,\\
A_{\nu_i}^{\mathrm{rev}}
&=
-\frac{1}{\gamma\tau}\Phi'(x_i)
+\frac{1}{\gamma\tau}F_i .
\end{align}

The irreversible probability current is therefore
\begin{equation}
J_{\nu_i}^{\mathrm{irr}}
=
A_{\nu_i}^{\mathrm{irr}} P - \frac{D}{\tau^2}\partial_{\nu_i}P
=
-\sum_j \Lambda_{ij}(\mathbf x)\nu_j\, P - \frac{D}{\tau^2}\partial_{\nu_i}P.
\end{equation}
We use the probability function calculated in \cite{Martin2021} in the equation above
\begin{widetext}
\begin{align}
P \sim e^{-\frac{\varphi + \sum_{i=1}^N \frac{\tau \nu_i^2}{2}}{D}} \left\{ 1 - \frac{\tau}{2D} \sum_{i=1}^N \left[ (\partial_i \varphi)^2 + \tau \sum_{j=1}^N \nu_i \nu_j \partial_i \partial_j \varphi - 3D \partial_i^2 \varphi \right] \right.\nonumber \\
\left. + \frac{\tau^2}{6D} \sum_{i=1}^N \nu_i \partial_i \sum_{j=1}^N \left[ \tau \sum_{k=1}^N \nu_j \nu_k \partial_j \partial_k - 3D \partial_j^2 \right] \varphi + O(\tau^2) \right\}.
\end{align}
\end{widetext}
\rev{The cancellation can be made explicit. Writing $P = P_0(1+\tau P_1+\dots)$ with $P_0\propto e^{-(\varphi+\sum_i\tau\nu_i^2/2)/D}$, the diffusive part of the irreversible current is $-\frac{D}{\tau^2}\partial_{\nu_i}P$. Differentiating the Gaussian velocity factor of $P_0$ gives $\partial_{\nu_i}P_0 = -\frac{\tau\nu_i}{D}P_0$, hence
\begin{equation}
-\frac{D}{\tau^2}\partial_{\nu_i}P_0 = \frac{\nu_i}{\tau}P_0 ,
\label{gaussianderivative}
\end{equation}
which cancels exactly the $\frac{1}{\tau}\delta_{ij}$ part of $-\sum_j\Lambda_{ij}\nu_jP$. Differentiating the $\mathcal O(\tau)$ bracket of $P$, whose velocity-dependent piece is $-\frac{\tau^2}{2D}\sum_{jk}\nu_j\nu_k\partial_j\partial_k\varphi$, gives a contribution $\frac{1}{\gamma}\sum_j\partial_i\partial_j\varphi\,\nu_jP$ at the same order in $\tau$, which cancels the remaining curvature part $\frac{1}{\gamma}\left[\Phi''(x_i)\delta_{ij}+\partial_{x_ix_j}V(x_i-x_j)\right]\nu_j$ of $\Lambda_{ij}$, since $\partial_i\partial_j\varphi$ is exactly that Hessian. All terms linear in $\boldsymbol\nu$ therefore cancel identically, and the first surviving contribution comes from the $\mathcal O(\tau^2)$ term of $P$, which upon differentiation produces the velocity-independent third-derivative term quoted below.}

$-\sum_j \Lambda_{ij}(\mathbf x)\nu_j$ in $J^{irr}$ is canceled by the exponential term in $P$ and the second term on the right hand side of the equation in $P$, leaving only the term
\begin{equation}
J_{\nu_i}^{\mathrm{irr}} = \frac{D}{2}\partial_{ijj}\varphi P + \mathcal{O}(\tau).
\end{equation}

The steady-state entropy production rate can be calculated from Spinney–Ford formula \cite{Spinney2012Entropy} with the diffusion matrix $(D^{(\nu)})^{-1}_{ij}=\frac{\tau^2}{D}\delta_{ij}$ 
\begin{widetext}
\begin{equation}
\dot S_{\mathrm{tot}}
=
\int d\mathbf x\, d\boldsymbol\nu\;
\sum_{i,j}
\frac{
J_{\nu_i}^{\mathrm{irr}}
(D^{(\nu)})^{-1}_{ij}
J_{\nu_j}^{\mathrm{irr}}
}{P}=
\frac{D\tau^2}{4}
\left\langle \left(\sum_{ij}\partial_{ijj}\varphi\right)^2
\right\rangle+\mathcal{O}(\tau^3).
\end{equation}
\end{widetext}

\begin{widetext}
\begin{equation}
\dot S_{\mathrm{tot}}
=
\frac{D\tau^2}{4}
\left\langle \left[\sum_{i}\left(\Phi^{\prime\prime}(x_i)+2\sum_{j\neq i}V^{\prime\prime}(x_i-x_j)\right)\right]^2
\right\rangle+\mathcal{O}(\tau^3).
\end{equation}
\end{widetext}

 The final entropy production under the mean field approximation reduces to
\begin{equation}
\dot S_{\mathrm{tot}} =
\frac{D\tau^2}{4}
\int dx \, \rho(x)\,
\left[\partial_x^3 \varphi_{\mathrm{eff}}(x)\right]^2,
\end{equation}
where the effective potential is given by
\begin{equation}
\varphi_{\mathrm{eff}}(x)
=
\Phi(x)
+ 2 \int dx^{\prime}\rho(x^{\prime})V(x-x^{\prime}).
\end{equation}

Similarly the medium entropy production is 
\begin{widetext}
\begin{align}
d\mathcal S_{\mathrm{med}} =
\sum_i \Bigg[
& \frac{\tau^2}{D}\left(-\sum_j M_{ij}\nu_j\right)\circ d\nu_i  - \frac{\tau^2}{D}
\left(A^{\mathrm{rev}}_{\nu_i} A^{\mathrm{ir}}_{\nu_i}\right) dt \nonumber\\
& + \frac{1}{p}\frac{\partial (A_i p)}{\partial \nu_i} dt
- \frac{D}{\tau^2}\frac{1}{p}\frac{\partial^2 p}{\partial \nu_i^2} dt
\Bigg]
\end{align}
\end{widetext}

Averaging this we obtain using $\langle\boldsymbol{\nu}^T\boldsymbol{\nu}\rangle = \frac{D}{\tau^2}\mathbf{M}^{-1}$
\begin{widetext}
\begin{equation}
\left\langle \dot{\mathcal S}_{\mathrm{med}} \right\rangle =
\frac{\tau^2}{D}
\sum_i
\left\langle
\left(\sum_j M_{ij}\nu_j\right)^2
\right\rangle 
=
\frac{\tau^2}{D}
\left\langle
\boldsymbol{\nu}^T \mathbf{M}^T \mathbf{M} \boldsymbol{\nu}
\right\rangle = \mathbf{M}
.
\end{equation}
\end{widetext}

\bibliography{bibliography}

\end{document}